\documentclass[preprint]{iucr}  
\RequirePackage{graphicx} 

     \journalcode{J}              
\usepackage{amsmath,amssymb}
\usepackage{xcolor}

\begin{document}  



\title{Elongated particles in flow: Commentary on small angle scattering investigations}


\author[a]{Guan-Rong}{Huang}
\author[b]{Lionel}{Porcar}
\author[c]{Ryan P.}{Murphy}
\author[d]{Yuya}{Shinohara}
\author[e]{Yangyang}{Wang}
\author[e]{Jan-Michael}{Carrillo}
\author[e]{Bobby G.}{Sumpter}
\author[f]{Chi-Huan}{Tung}
\author[f]{Changwoo}{Do}
\cauthor[f]{Wei-Ren}{Chen}{chenw@ornl.gov}

\aff[a]{Department of Engineering and System Science, National Tsing Hua University, Hsinchu 300044, \country{Taiwan}}
\aff[b]{Institut Laue-Langevin, B.P. 156, F-38042 Grenoble Cedex 9, \country{France}}
\aff[c]{NIST Center for Neutron Research,  National Institute of Standards and Technology, Gaithersburg, MD 20878, \country{United States}}
\aff[d]{Materials Science and Technology Division, Oak Ridge National Laboratory, Oak Ridge, TN 37831, \country{United States}}
\aff[e]{Center for Nanophase Materials Sciences, Oak Ridge National Laboratory, Oak Ridge, TN 37831, \country{United States}}
\aff[f]{Neutron Scattering Division, Oak Ridge National Laboratory, Oak Ridge, TN 37831, \country{United States}}


\maketitle   

\begin{synopsis}
We critically examine mathematical tools to invert the orientation distribution of flowing elongated objects from anisotropic small angle scattering data. This evaluation aims to advance understanding of the interplay between flow dynamics and object orientation, benefiting fluid dynamics and materials science.
\end{synopsis}

\clearpage

\begin{abstract}

This work thoroughly examines several analytical tools, each possessing a different level of mathematical intricacy, for the purpose of characterizing the orientation distribution function of elongated objects under flow. Our investigation places an emphasis on connecting the orientation distribution to the small angle scattering spectra measured experimentally. The diverse range of mathematical approaches investigated herein provides insights into the flow behavior of elongated particles from different perspectives and serves as powerful tools for elucidating the complex interplay between flow dynamics and the orientation distribution function.

\end{abstract}

\clearpage


\section{Introduction}

In recent decades, there has been significant interest in understanding the dynamic behavior and alignment of flowing objects, driven by both scientific curiosity and the demands of advancing technology. Theoretical descriptions of these systems have become a prominent research topic, as the complex interplay among hydrodynamic forces, Brownian motion, and direct inter-particle interactions profoundly influences the orientational ordering of elongated particles across diverse flow regimes. 

Understanding the flow behavior of rod-like particles is particularly relevant in numerous industrial sectors, where manipulating anisotropic macromolecules is essential for creating tailored products with specific characteristics. Such materials include, but are not limited to, stiff polymers \cite{Gantenbein2018, Fratzl2007}, nanostructured films \cite{Long2012, Blell2017}, fibers \cite{Hakansson2014, Richard2013}, and non-spherical nanoparticles \cite{Patil2010, Wu2017}, all of which can be strategically induced into a state of flow to confer desired properties such as high modulus or specific geometries.

Understanding the structure-property relationships of these materials during flow represents a key focus in their processing. Building upon Onsager's pioneering work in density functional theory calculations \cite{Onsager}, subsequent theoretical inquiries have concentrated on revealing the phase behaviors of various soft matter systems containing aligned elongated components \cite{MS1958, HJL1972, Faber1972, Stephen1974, Singh1991, Lekkerkerker1992, Singh2000, Stokes2020}. Further studies have adopted sophisticated mathematical tools such as constitutive modeling and the application of the Smoluchowski equations \cite{Jeffery1922, Batchelor1970, Hinch1976, Bird1987, Doi1986, Larson2013, Hakansson2016} to delve deeper into the behavior of rod suspensions under flow and deformation conditions.

Central to this theoretical and computational research landscape is the emergence of the orientation distribution function (ODF) as a pivotal metric of interest. The ODF acts as a detailed descriptor in the Hamiltonian of systems consisting of elongated particles \cite{Onsager}, accounting for the complex impacts of orientation entropy, inter-rod exclusion effects, and long-range interactions on the phase behavior and rheological properties of rod suspensions. Moreover, its importance lies in its direct correlation with significant mechanical properties such as shear and hydrodynamic stresses. As a result, the ODF plays a fundamental role in characterizing the properties and alignment of rod suspensions, offering invaluable insights through the perspective of statistical mechanics.

In this commentary, we explore small angle neutron scattering, specifically examining its contemporary applications in understanding the spatial organization of elongated objects amidst diverse flow conditions. Our emphasis lies in a thorough analysis of scattering spectra, represented by two-point spatial correlation functions in reciprocal space. The inherent angular anisotropy within these functions conceals information regarding the ODF of elongated objects in non-equilibrium states.

The primary objective of this feature article is to comprehensively investigate the historical evolution of small angle scattering data analysis in flowing materials. Existing literature extensively covers the broader application of small angle neutron scattering (SANS) techniques in rheology, often referred to as rheo-SANS, and provides detailed insights into associated sample environments \cite{Lindner1991, Lindner2002, Lindner2011, Martel2021, Gilbert2024}. However, the need for this literature review became apparent as the authors encountered difficulties navigating the existing body of work on mathematical strategies for reconstructing distorted structures of deforming materials based on the anisotropy of small angle scattering. This work uniquely focuses on exploring tailored mathematical strategies for inverting the orientation of elongated objects in motion from their small angle scattering. Throughout our discourse, we compare theoretical predictions with computational results and experimental scattering data, enriching the understanding of the subject matter and opening avenues for new insights in this complex field.

\section{Heuristic Considerations on Inverting Flow-induced Alignment from Spectral Anisotropy}

Before delving into the historical evolution of methods for extracting the ODF from scattering data, it is instructive to provide a mathematical description of the spectral inversion process. A logical starting point for this explanation is the rotation matrix \cite{Goldstein}, which determines the new coordinates of a vector after rotation. The alignment of elongated particles in motion is significantly influenced by the interplay of thermodynamic and fluid mechanical forces. In a quiescent solution, particles assume random orientations primarily due to the thermodynamic principle of entropy. However, when subjected to mechanical forces, such as in the case of a sheared solution, the flow field disrupts the equilibrium of particle orientation. This disruption initiates a complex interplay between thermodynamic forces and fluid mechanical dynamics, ultimately leading to particle rotation. To grasp this concept within the framework of the rotation matrix, it is essential to establish the degree of freedom required to uniquely determine the rotational direction in real space.

\begin{figure}
\centerline{
  \includegraphics[width=\linewidth]{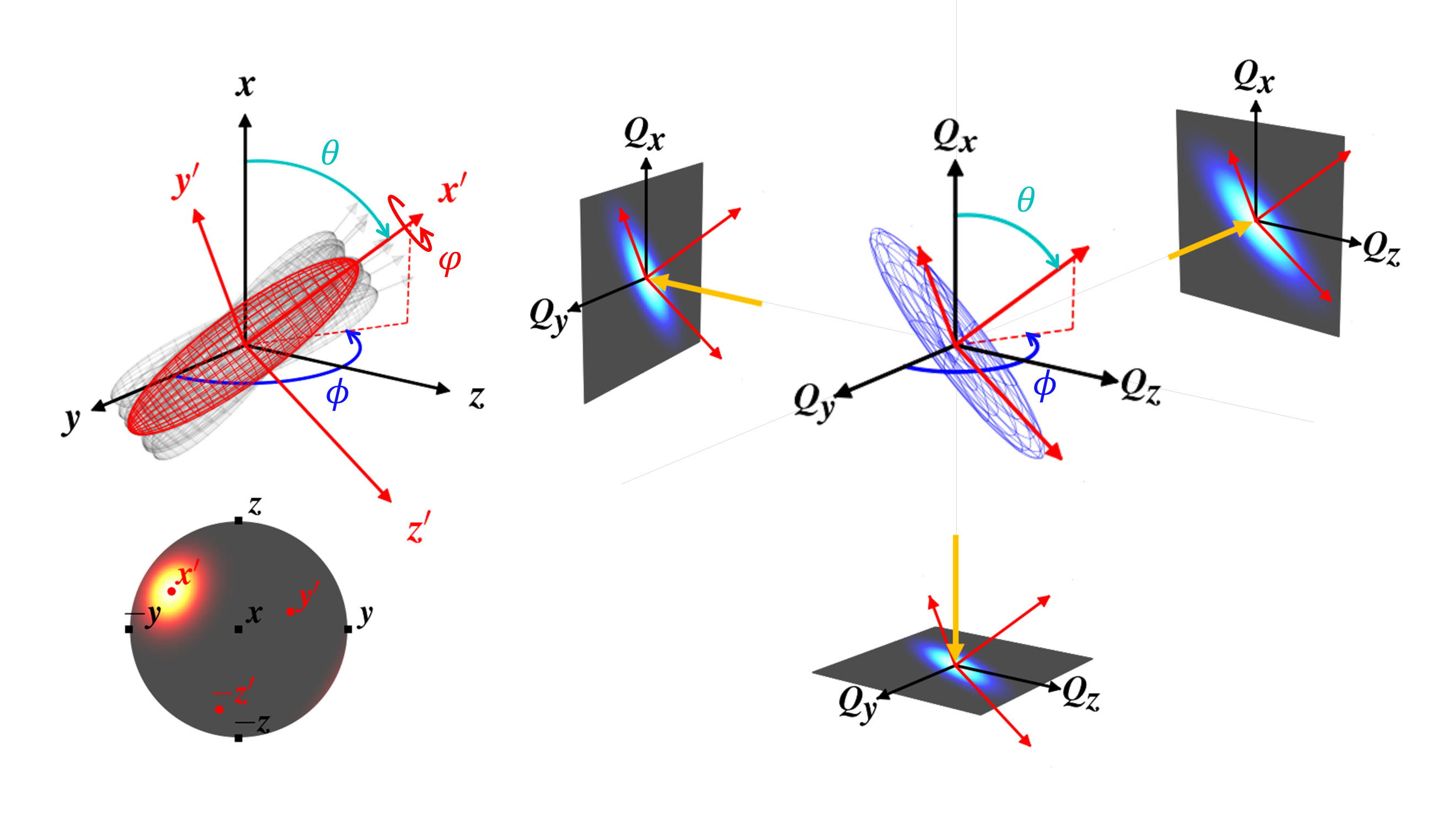}
}  
\caption{A schematic representation illustrating mechanically-driven flowing objects (red ellipsoid) and their corresponding scattering intensity in reciprocal space. The black arrows denote the laboratory reference coordinate system. The red arrows depict the coordinate system associated with the tagged particle. The yellow arrows indicate the directions of incident radiation probing the spatial correlation of flowing particles. Due to the presence of preferred orientation among the flowing particles, the scattering intensity $I(\textbf{Q})$, with one of its intensity isosurfaces represented by the blue ellipsoid, exhibits angular anisotropy. The inset in the bottom left corner depicts spherical probability density functions, illustrating the orientation of flowing particles. Black letters indicate the laboratory coordinate system, while red letters denote the system associated with the tagged flowing particle.
}  
\label{fig:1}
\end{figure}

When rigid particles undergo a rotation operation, their relative positions and distances are invariant, a fundamental property known as an isometric transformation \cite{Goldstein}. Describing this transformation involves introducing a rotation matrix denoted as \(\mathbb{R}\), which illustrates how coordinates or vectors evolve under this operation \cite{Goldstein}. A key feature of rotation matrices is characterized by orthonormality for preserving distances and angles. \(\mathbb{R}\) in three-dimensional space comprises a $3 \times 3$ array consisting of 9 elements, typically embodying 9 degrees of freedom. Given that \(\mathbb{R}\) is an orthonormal matrix, signifying that its columns are unit vectors, it imposes a restriction, thereby diminishing the degrees of freedom by one for each column. Hence, starting with an initial 9 degrees of freedom, we deduct 3, resulting in 6. Furthermore, the orthonormal nature entails that all columns are orthogonal to each other. Consequently, for every pair of columns, one degree of freedom is relinquished, totaling $\frac{3(3-1)}{2} = 3$ degrees of freedom. Hence, the net degree of freedom becomes $6-3 = 3$. According to Euler’s rotation theorem \cite{Euler}, any rotation can be described using three parameters. Based on this mathematical framework, the orientation of rotated pole vector $\textbf{x}^\prime$ relative to the original $\textbf{x}$ can be determined by the azimuthal angle $\phi$ and polar angle $\theta$ in the spherical coordinates. $\varphi$ characterizes the degree of freedom associated with the spin of the tagged particle relative to the major axis. Fig.~\ref{fig:1} illustrates the relationship among these angles in the physics convention.

The rotation from $\textbf{x}$ to $\textbf{x}^\prime$ can be succinctly captured through the Rodrigues' rotation formula as \cite{Morawiec}:
\begin{equation}
    \mathbb{R} = \mathbb{I}+\mathbb{K}+\frac{1-\cos{\theta}}{\sin^2{\theta}}\mathbb{K}^2,
\label{eq:2.0}
\end{equation}
where $\mathbb{I}$ is the unit matrix, $\mathbb{K}$ is a skew-symmetric matrix defined as
\begin{equation}
    \begin{bmatrix}
        0 & -k_z & k_y\\
        k_z & 0 & -k_x\\
        -k_y & k_x & 0
    \end{bmatrix}.
\end{equation}
The variables $k_x$, $k_y$, and $k_z$ represent the components of $\mathbf{k}=\mathbf{x}\times\mathbf{x}^\prime$. Additionally, $\varphi$ is necessary to describe the intrinsic rotation about $\mathbf{x}^\prime$. Drawing from the mathematical framework of the rotation matrix, the procedure for deriving the ODF for elongated particles in motion via small angle scattering is outlined in Fig.~\ref{fig:1}. The present discussion focuses on dilute solutions of particles where inter-particle interactions are negligible. Within this simplified context, we aggregate the center-of-mass of each individual particle, depicted as gray ellipsoids, to that of the tagged red particle. The physical significance of the ODF becomes apparent upon examination of the sketch containing the red ellipsoid in Fig.~\ref{fig:1}: It offers a mathematical depiction of how a group of flowing particles aligns around the averaged orientation $\textbf{x}^\prime$. This alignment is represented by the orientation of gray arrows, collectively indicating a certain variance and some deviation from $\textbf{x}^\prime$.

This alignment leads to anisotropic angular distributions of three-dimensional scattering intensity $I(\textbf{Q})$ in reciprocal space, where $\textbf{Q}$ represents the momentum transfer between the incident and scattered beam. The blue ellipsoid in Fig.~\ref{fig:1} represents one of its intensity isosurfaces. Depending on the direction of incident radiation (the yellow arrows) --- either neutrons or X-rays --- the projections of scattering intensities on different planes display varying degrees of angular anisotropy.

To implement a reliable and practical ODF inversion, various boundary conditions associated with this mathematical problem need to be addressed. The first constraint is inherent in the scattering technique. In scattering experiments, structural information is embedded within a mathematical framework of two-point static correlation \cite{Marshall}. Consequently, certain structural details, such as higher-order spatial correlations, are expected to be lost in this coarse-grained representation. Since the collected anisotropic scattering intensity encompasses contributions from both intra-particle spatial correlation and the ODF of the flowing particle assembly, it is crucial to specify the particle shape \textit{a priori}. This specification helps disentangle the covariance between the ODF and the particle form factor during regression analysis, thereby ensuring the unique determination of the ODF from the measured spectra.

Additionally, determining the exact number of flow planes to be probed is crucial for providing sufficient informational inputs to ascertain the ODF. This determination hinges on the orientation of the $\textbf{x}^\prime$ direction relative to the laboratory reference frame, which is collectively determined by the deformation gradient of the applied field and the axial symmetry of the particle geometry.

Therefore, the problem to be addressed involves the inversion of the three-dimensional ODF in real space from the experimentally measured two-dimensional angular anisotropy of the small angle scattering intensity in reciprocal $\textbf{Q}$ space, expressed as static two-point correlation. These endeavors must be undertaken while adhering to the mathematical constraints outlined earlier. 

\section{Historical Survey of the Mathematical Methodologies in Data Analysis}\label{Sec:II}

In this section, we review the current spectral inversion approaches. Our objective, backed by comparative evaluation and analysis, is to shed light on the inherent strengths and weaknesses of each approach, thereby contributing to a thorough comprehension of spectral inversion methodologies. It is worth noting that initially, various spectral analysis algorithms may have been developed to address different aspects of structural distortion in flowing objects based on their scattering features. In this report, our main emphasis is on assessing their capability to invert the ODF quantitatively.

The anisotropic small angle scattering cross section of flowing elongated objects can be succinctly expressed without loss of generality by the following equation:
\begin{equation}
I(\textbf{Q}) = \int \mathrm{d}\Omega P(\textbf{Q}, \Omega) f(\Omega).
\label{eq:2.1}
\end{equation}
Here, $\Omega$ and $\mathrm{d}\Omega$ respectively denote the orientation and the differential solid angle, $\mathrm{d}\phi\mathrm{d}\theta\sin{\theta}$ in three-dimensional real space, while $P(\textbf{Q}, \Omega)$ is the intra-particle structure factor associated with a specific orientation defined by $\Omega$, and $f(\Omega)$ represents the ODF. It is crucial to note that $I(\textbf{Q})$ is a three-dimensional quantity in reciprocal space. 

Within the mathematical framework of Eqn.~\eqref{eq:2.1}, inferring $f(\Omega)$ from the inversion of $I(\textbf{Q})$ can be approached in two distinct manners. When a predefined model of $f(\Omega)$ is available, a parametric approach is feasible. When such information is unavailable, a non-parametric approach becomes necessary. An essential aspect of this structural inversion problem is to ascertain whether a provided parametric model of $f(\Omega)$ is sufficiently sophisticated to describe the statistical characteristics of the ODF, or if the adoption of non-parametric modeling becomes necessary.

\subsection{Parametric Regression Analysis}

When one holds \textit{a priori} confidence in the accuracy of the analytical expression of the input $f(\Omega)$, a parametric regression analysis can be used. The process comprises two distinct steps: firstly, identifying an appropriate analytical formulation of $f(\Omega)$ for pertinent pre-selected parameters governing the orientation distribution of flowing particles, and subsequently, determining the numerical values of these parameters through regression analysis techniques such as least squares or maximum likelihood methods applied to the observed scattering intensity, which is the projection of $I(\textbf{Q})$ onto a certain plane. 

The advantage of this parametric regression is evident: by expressing the ODF analytically, comprehension and interpretation become more accessible. Over recent decades, numerous functional forms of $f(\Omega)$ have been proposed to facilitate spectral inversion analysis. These models, though not exhaustively listed, commonly include simple one-dimensional functions of the polar angle $f(\theta)$, addressing spectral anisotropy characterized by varying degrees of axial symmetry. Examples encompass the Boltzmann function \cite{Foerster2005, Helgeson2024}, Gaussian function \cite{Odijk1986, Zhou2004, Wang2006, Ch2011, Tabor2022}, Generalized normal distribution \cite{Vainio2014}, Lorentzian distribution \cite{Hwang2000, Wang2006, Ch2011}, Heaviside function \cite{Foerster2005}, Maier-Saupe function \cite{MS1958,Maier1959,Maier1960,Luckhurst1977, De1993,Lemaire2002,Porcar2008,Lang2016, Lang2019}, Onsager's trial function \cite{Onsager, Wu2014, Porcar2008, Franco2008, Franco2009}, and Oldenbourg function \cite{Caspar1988, Caspar1993}. These functions exhibit distinctive traits ranging from abrupt step-function transitions to much slower exponential decay. Alternatively, more sophisticated models express dependencies on both polar and azimuthal angles of $f(\Omega)$ \cite{Scheraga1951, Jerrard1959, Okagawa1973, Hayter1984, Doi1986, Herbst1986, Hayter1987, Kalus1988, Penfold1988, Penfold1991, Bihannic2010, Bihannic2011, Helgeson2024}. For a comprehensive understanding of the mathematical formulations underlying these ODF models, readers are encouraged to consult relevant citations provided above.

While it may offer mathematical convenience, selecting a particular functional form inherently imposes limitations on the parametric approach. It is widely acknowledged in the existing literature that the precise analytical expression of $f(\Omega)$ for various flowing objects is typically unknown \textit{a} \textit{priori}. In situations involving highly oriented elongated particles, the choice of a specific functional expression may not be critical, as the axial symmetry of the ODF can often be effectively approximated by a Gaussian function with a narrow variance. However, in cases of moderately oriented systems, a trial-and-error approach frequently becomes necessary in regression analysis to ascertain the most appropriate expression for the ODF. This iterative process aims to accurately capture the unique characteristics of spectral anisotropy by minimizing the squared deviation between the scattering data and the model function $f(\Omega)$.

Nevertheless, the final determination of $f(\Omega)$ cannot rely only on the level of quantitative agreement between the optimized theoretical model and the experimental data. For instance, in a SANS study of sheared worm-like micellar solutions \cite{Foerster2005}, it was found that Onsager's trial function offered the closest match between calculated and experimental scattering patterns. This result suggests that the systems can be treated as dilute solutions, because the theoretical foundation of Onsager's trial function is grounded in the central limit theorem \cite{Kardar}, which implies that the orientational alignment of each elongated particle should be independent of neighboring particles. However, further scrutiny through complementary nonlinear rheological assessments revealed the highly interactive nature of the same system, as evidenced by Einstein's law \cite{Oswald}. This observation presents a contrasting perspective, indicating that the system exhibits strong collective behavior. 

This example underscores a fundamental limitation of parametric spectral analysis in extracting $f(\Omega)$: when the formulation is known and accurate, the chosen parametric model can aptly capture the underlying $f(\Omega)$. However, opting for an incorrect functional form can lead to a greater bias compared to other complementary analyses.

\subsection{Non-parametric Regression Analysis}

Non-parametric regression methods, unlike their parametric counterparts, refrain from assuming a specific mathematical form for $f(\Omega)$. This characteristic provides them with the flexibility to discern any underlying functional form within experimental scattering data. One prominent approach in non-parametric regression involves basis expansion. In this technique, the function space of $f(\Omega)$ is expanded using carefully selected orthonormal basis functions. The basis functions act as fundamental units, proficient at capturing complex relationships within the data without imposing predefined assumptions about the form of the ODF. As a result, this method enables the data-driven shaping of relationships through the selection of basis functions, freeing the analysis from predetermined model structures.

\subsubsection{Criteria for Selecting Basis Functions in Spectral Analysis of Flowing Materials}

A natural question to ask is, among the available orthogonal basis sets \cite{Roman, Arfken, Strang}, which one is most suited for the inversion of the ODF derived from spectral anisotropy?

In the context of deriving $f(\Omega)$ from $I(\textbf{Q})$ for elongated objects under flow, the orthonormal set of real spherical harmonics, denoted as $Y_l^m$ where $l$ represents the degree and $m$ the order, is the optimal choice for two key reasons. The first reason is that this scattering study focuses on understanding the rotation of elongated objects in varying flow conditions, as depicted in Fig.~\ref{fig:1}. Therefore, selecting $Y_l^m$ is natural because it is compatible with boundary conditions and spatial symmetry and is well-established as eigenfunctions of rotation operators along arbitrary directions \cite{Schiff, Varshalovich1988}. The set ${Y_l^m}$ constitutes a complete basis and can effectively expand any angular function of $\Omega$ with boundary conditions consistent in three-dimensional space \cite{Arfken}. Moreover, when employing the spherical harmonic expansion (SHE) for $f(\Omega)$ and $I(\textbf{Q})$, its orthogonality and completeness ensures that the expansion coefficients of an angular distribution are both unique and independent. The inherent spatial periodicity in angular variables is naturally accommodated within the $Y_l^m$ basis. Furthermore, these harmonics have been extensively used to describe the angular distribution of particles within a spherically symmetric field with specific orbital angular momentum $l$ and projection $m$. Thus, their parity inherently aligns with the anisotropic symmetry observed in both $f(\Omega)$ and $I(\textbf{Q})$. 

The second merit of $Y_l^m$ is based on the scattering technique itself: $I(\mathbf{Q})$ is related to the Fourier transform of the scattering length density profile of particles. This profile represents the superposition of plane waves $\exp{(-i\mathbf{Q} \cdot \mathbf{r})}$ that contain structural information at various length scales. The expression for $\exp{(-i\mathbf{Q} \cdot \mathbf{r})}$ can be expanded as follows \cite{Varshalovich1988,Arfken,HuangJPCL}:
\begin{eqnarray}
    \exp{(-i\mathbf{Q} \cdot \mathbf{r})} = \sum_{l=0}^{\infty}\sum_{m=-l}^{l} (-i)^l j_{l}(Qr)Y_l^m(\Omega)Y_l^m(\hat{\mathbf{Q}}), 
    \label{eq:3a1}
\end{eqnarray}
where $j_l(Qr)$ are spherical Bessel functions of order $l$, $Q$ and $r$ are respectively the magnitudes of $\mathbf{Q}$ and $\mathbf{r}$. $\Omega$ and $\hat{\mathbf{Q}}$ are the orientations of $\mathbf{r}$ and $\mathbf{Q}$ specified by the polar angle $\theta$ and azimuthal angle $\phi$ in real and reciprocal space. Specifically,
\begin{equation}
\mathrm{d}^3\mathbf{Q} = \mathrm{d}\hat{\mathbf{Q}}\mathrm{d}QQ^2 = \mathrm{d}\phi \mathrm{d}\theta \mathrm{d}Q Q^2\sin{\theta},
\label{eq:3.a2}
\end{equation}
and
\begin{equation}
\mathrm{d}^3\mathbf{r} = \mathrm{d}\Omega\mathrm{d}rr^2 =  \mathrm{d}\phi \mathrm{d}\theta \mathrm{d}r r^2 \sin{\theta}.
\label{eq:3.a3}
\end{equation}
By expanding $\exp{(-i\mathbf{Q} \cdot \mathbf{r})}$ using spherical harmonics $Y_l^m$, it becomes evident that the real-space angular component associated with the ODF $Y_l^m(\Omega)$ and the reciprocal-space $\hat{\mathbf{Q}}$ component $Y_l^m(\hat{\mathbf{Q}})$ are decoupled, as demonstrated by Eqn. \eqref{eq:3a1}. Leveraging the spherical harmonic addition theorem \cite{Arfken}, the term $Y_l^m(\Omega)Y_l^m(\hat{\mathbf{Q}})$ in Eqn. \eqref{eq:3a1} elucidates the angle between $\mathbf{r}$ and $\mathbf{Q}$ within the phase of the plane wave. The rotationally-invariant inner product between $\mathbf{Q}$ and $\mathbf{r}$ is preserved in the SHE for different reference frames. Consequently, using $Y_l^m$ as the fundamental basis allows for the generation of specific geometrical symmetries and boundary conditions, facilitating the separation of variables $\Omega$ and $\hat{\mathbf{Q}}$ during the Fourier transform process. This separation greatly simplifies the inversion of the ODF of flowing objects from the corresponding $I(\textbf{Q})$. 

Before delving into the mathematical framework of the SHE, it is essential to address a fundamental difference in its application for structural analysis in simulation versus scattering experiments: In simulations, the spatial arrangements of particles in deforming and flowing materials are typically represented by computationally-generated trajectories in three-dimensional real space. With the coordinates of each particle available, analyzing structural distortion using SHE is mathematically straightforward. However, in practical scattering experiments, only conditional information is available. This information is typically in the form of a two-dimensional projection of $I(\textbf{Q})$ from the probed flow plane. Additionally, the orthogonality of spherical harmonic basis vectors changes from three-dimensional to two-dimensional space in this scenario. These limitations must be carefully considered when reconstructing the ODF of flowing elongated particles from their scattering data.

\subsubsection{General Mathematical Framework}

Let us outline the SHE application for extracting the ODF from $I(\textbf{Q})$. Initially, we will represent the $I(\textbf{Q})$ in terms of spherical harmonic basis functions as follows: 
\begin{equation}
I(\textbf{Q}) = \sum_{l=0}^{\infty} \sum_{m = -l}^{l}I_l^m(Q)Y_l^m(\hat{\textbf{Q}}),
\label{eq:3.a}
\end{equation}
where $\hat{\textbf{Q}}$ is the orientation along $\textbf{Q}$. $Y_l^m(\hat{\textbf{Q}})$ denotes the Real Spherical Harmonics (RSH) of degree $l$ and order $m$ with argument $\hat{\textbf{Q}}$. These functions satisfy the following orthogonality condition:
\begin{equation}
\int \mathrm{d}\hat{\textbf{Q}} Y_l^m(\hat{\textbf{Q}}) Y_{l'}^{m'} (\hat{\textbf{Q}}) = 4 \pi \delta_{l,l'} \delta_{m,m'},
\label{eq:3.a1}
\end{equation}
where $\mathrm{d}\hat{\textbf{Q}}$ represents the differential solid angle: $\mathrm{d}\phi\mathrm{d}\theta \sin{\theta}$ in three-dimensional $\textbf{Q}$ space. Here, $\delta_{l,l'}$ and $\delta_{m,m'}$ are Kronecker delta functions, equal to one if the subscripts are identical and zero otherwise. Furthermore, $I_l^m(Q)$ are the the coefficients corresponding to each $Y_l^m(\hat{\textbf{Q}})$, where $Q \equiv |\textbf{Q}|$. 
Likewise, the ODF $f(\Omega)$ can be expanded as:
\begin{equation}
f(\Omega) = \sum_{l=0}^{\infty} \sum_{m = -l}^{l} S_l^m Y_l^m(\Omega),
\label{eq:3.b}
\end{equation}
The coefficients $S_l^m$ function as weighting factors that quantify the inherent spatial arrangement characteristic of each $Y_l^m (\Omega)$. Because each basis function $Y_l^m (\Omega)$ embodies distinct spatial symmetry features, the numerical values of $S_l^m$ precisely capture the extent of spatial ordering corresponding to each $Y_l^m (\Omega)$. Consequently, $S_l^m$ commonly earns the designation of \textit{order parameters}.

In order to derive \( f(\Omega) \) from \( I(\mathbf{Q}) \), it is necessary to partition the intra-particle spatial correlation, denoted as \( P(\mathbf{Q}) \), in Eq.~\eqref{eq:2.1}. \( P(\mathbf{Q}) \) can be expressed by the following equation:
\begin{equation}
P(\textbf{Q}) = \int \mathrm{d}\Omega f(\Omega) P(\textbf{Q}, \Omega).
\label{eq:3.c}
\end{equation}
According to Eqn.~\eqref{eq:3.c}, $P(\textbf{Q})$ represents the angularly averaged intra-particle spatial correlation function given $f(\Omega)$ as the ODF. It is essential to note that if the set $f(\Omega)$ follows a continuous uniform distribution across $\Omega$, indicating random particle orientations, then $P(\textbf{Q})$ will demonstrate isotropy in three-dimensional reciprocal $\textbf{Q}$ space. This implies it can be expressed as $P(|\textbf{Q}|)$. As illustrated in Fig.~\ref{fig:1}, any two-dimensional projection of this angularly isotropic $P(|\textbf{Q}|)$ will also exhibit isotropy, resembling the familiar particle form factor $P(Q)$.

Using SHE, $P(\textbf{Q})$ in Eqn.~\eqref{eq:3.c} can be reformulated as
\begin{equation}
P(\textbf{Q}) = \sum_{l=0}^{\infty} \sum_{m = -l}^{l}S_l^m P_l^m(Q) Y_l^m (\hat{\textbf{Q}}),
\label{eq:3.d}
\end{equation}
where
\begin{equation}
P_l^m(Q) = \frac{1}{16 \pi ^2} \int \mathrm{d}\hat{\textbf{Q}} \mathrm{d}\Omega P(\textbf{Q}, \Omega) Y_l^m(\Omega) Y_l^m(\hat{\textbf{Q}}).
\label{eq:3.e}
\end{equation}
To facilitate the extraction of $f(\Omega)$, let's further define the following quantity from $I(\textbf{Q})$:
\begin{equation}
\hat{S}_l^m(Q) = \frac{\int \mathrm{d}\hat{\textbf{Q}} I(\textbf{Q}) Y_l^m (\hat{\textbf{Q}})}{\int \mathrm{d}\hat{\textbf{Q}} I(\textbf{Q}) Y_0^0 (\hat{\textbf{Q}})}.
\label{eq:3.f}
\end{equation}
In this context, $\hat{S}_l^m(Q)$ is termed the \textit{scattering order parameter}, serving as a qualitative indicator of spectral anisotropy within the framework of spherical harmonic functions. Upon substituting Eqn.~\eqref{eq:3.a} into Eqn.~\eqref{eq:3.f} and using the orthogonality of spherical harmonic functions, we note that:
\begin{equation}
\hat{S}_l^m(Q) = \frac{I_l^m(Q)}{I_0^0(Q)}.
\label{eq:3.g}
\end{equation}
In dilute solutions where inter-particle spatial correlation is negligible, the $I(\textbf{Q})$ corresponds to $P(\textbf{Q})$. Consequently, by Eqns.~\eqref{eq:3.d} and \eqref{eq:3.g}, we derive:
\begin{equation}
\hat{S}_l^m(Q) = \frac{P_l^m(Q)}{P_0^0(Q)}\frac{S_l^m}{S_0^0} = \frac{P_l^m(Q)}{P_0^0(Q)}S_l^m.
\label{eq:3.h}
\end{equation}
It's important to note that thus far in our discussion, we have focused solely on non-interacting systems, where the inter-particle interaction can be considered negligible. In concentrated systems, the corresponding scattering intensity in the high $Q$ limit primarily arises from intra-particle correlations, as the influence of inter-particle spatial correlations is minimal. Therefore, Eqn.~\eqref{eq:3.h} continues to be applicable for interacting systems.

The desired quantity $S_l^m$ can be computed using the following expression:
\begin{equation}
S_l^m = \frac{P_0^0(Q)}{P_l^m(Q)}\hat{S}_l^m(Q).
\label{eq:3.i}
\end{equation}
As $P_0^0(Q)$, $P_l^m(Q)$, and $\hat{S}_l^m(Q)$ can all be derived from $I(\textbf{Q})$, the ODF $f(\Omega)$ can be determined accordingly using Eqn.~\eqref{eq:3.b}. 

Before concluding the discussion in this subsection, it is crucial to emphasize that $I(\textbf{Q})$ exhibits even parity when the excess scattering length density of suspending particles is real and Friedel's law \cite{Friedel1913} is assumed to be valid. This inherent property of $I(\textbf{Q})$ imposes a constraint on spectral expansion using spherical harmonic functions. Specifically, only spherical harmonics $S_l^m$ with even integers for $l$ can be sampled in scattering experiments.

\subsubsection{Factors to Consider for Practical Spectral Analysis}

The preceding subsection has substantiated the viability of using SHE for determining the ODF of flowing elongated objects. However, the practical spectral analysis of real experimental data demands careful consideration of several critical factors. It is important to reiterate that $I(\textbf{Q})$, as defined in Eqns.~\eqref{eq:2.1} and \eqref{eq:3.a}, constitutes a \textit{three-dimensional} quantity. The experimental observable involves the projection of $I(\textbf{Q})$ onto specific flow planes, as illustrated by the Panels (d-f) in Fig.~\ref{fig:2}, which are \textit{two-dimensional} quantities. Hereafter, we delineate the strategy for extracting $f(\Omega)$ from these two-dimensional projections of $I(\textbf{Q})$.

\subsubsection{Spectra with Axial Symmetry}

\begin{figure}
\centerline{
  \includegraphics[width=\linewidth]{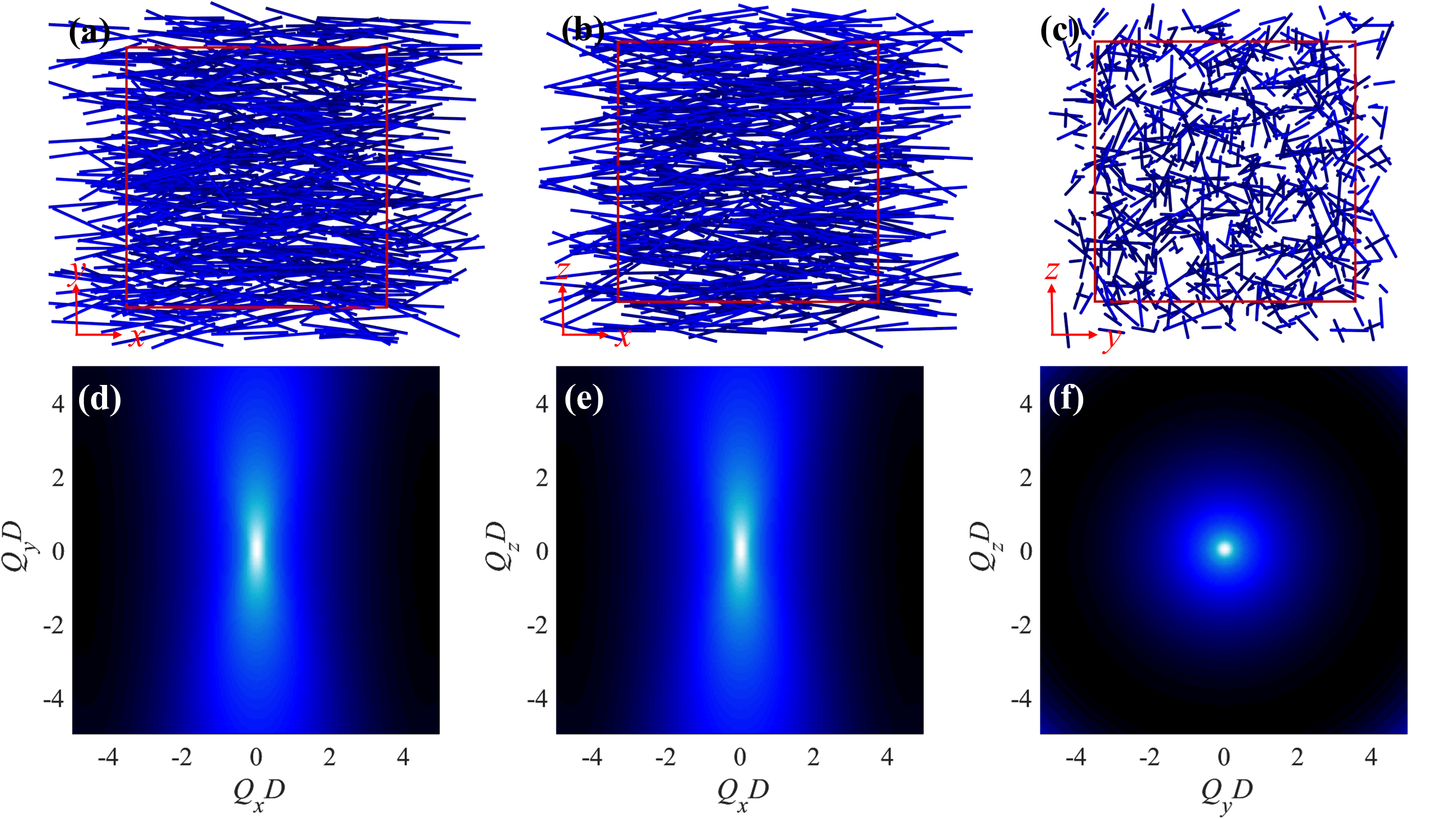}
}  
\caption{Snapshot of a sheared semi-dilute solution containing rigid rods with diameter $D$ and length $25D$. Real space projections from flow-velocity gradient ($x-y$), flow-vorticity ($x-z$), and velocity gradient-vorticity ($y-z$) planes are depicted in (a-c), respectively. Corresponding scattering spectra for these projections are shown in (d-f).
}  
\label{fig:2}
\end{figure}

Evidenced by existing experimental results, projections of $I(\textbf{Q})$ for numerous soft materials under uniaxial tension, Poiseuille flow, or worm-like micelles under Couette flow often display axial symmetry, as seen in Fig.~\ref{fig:2}(d-e), where the scattering intensity exhibits mirror symmetry about the vertical and horizontal axes. The spatial symmetry of spherical harmonic functions suggests that the reflection symmetry requires only basis functions where $m=0$ to be adequate for spectral analysis \cite{Huang2019}.

Consequently, Eqn.~\eqref{eq:3.d} can be simplified as:
\begin{equation}
I(\textbf{Q}) = \sum_{l=0,2,4,\cdots}^{\infty} S_l^0 P_l^0(Q) Y_l^0(\cos \theta).
\label{eq:3.i1}
\end{equation}
To test the validity of this approach, dissipative particle dynamics (DPD) simulations \cite{DPD} consisting of 216 rigid rods in a semi-dilute solution under shear were carried out. From this simple shear flow computational experiment, the projections of $I(\textbf{Q})$ from the flow velocity gradient, flow vorticity, and velocity gradient vorticity were sampled. Conventionally, these planes are identified as the $xy$-, $xz$-, and $yz$-planes. The intensity on the $xy$-plane, denoted as $I_{xy}(\textbf{Q})$, can be expressed as:
\begin{equation}
I_{xy}(\textbf{Q}) = \sum_{l=0,2,4,\cdots}^{\infty}S_l^0P_l^0(Q)Y_l^0(\cos\theta).
\label{eq:3.i2}
\end{equation}
Likewise,
\begin{equation}
I_{xz}(\textbf{Q}) = \sum_{l=0,2,4,\cdots}^{\infty}S_l^0P_l^0(Q)Y_l^0(\cos \theta),
\label{eq:3.i3}
\end{equation}
and
\begin{eqnarray}
I_{yz}(\textbf{Q}) &=&
\sum_{l=0,2,4,\cdots}^{\infty}S_l^0P_l^0(Q)Y_l^0(\cos{\frac{\pi}{2}}).
\label{eq:3.i4}
\end{eqnarray}
Eqn.~\eqref{eq:3.i4} suggests that $I_{yz}(\textbf{Q})$ demonstrates angular isotropy, as illustrated in Fig.~\ref{fig:2}(f). As a result, $I_{yz}(\textbf{Q})$ is of least significance for extracting $f(\Omega)$. 

\begin{figure}
\centerline{
  \includegraphics[width=\linewidth]{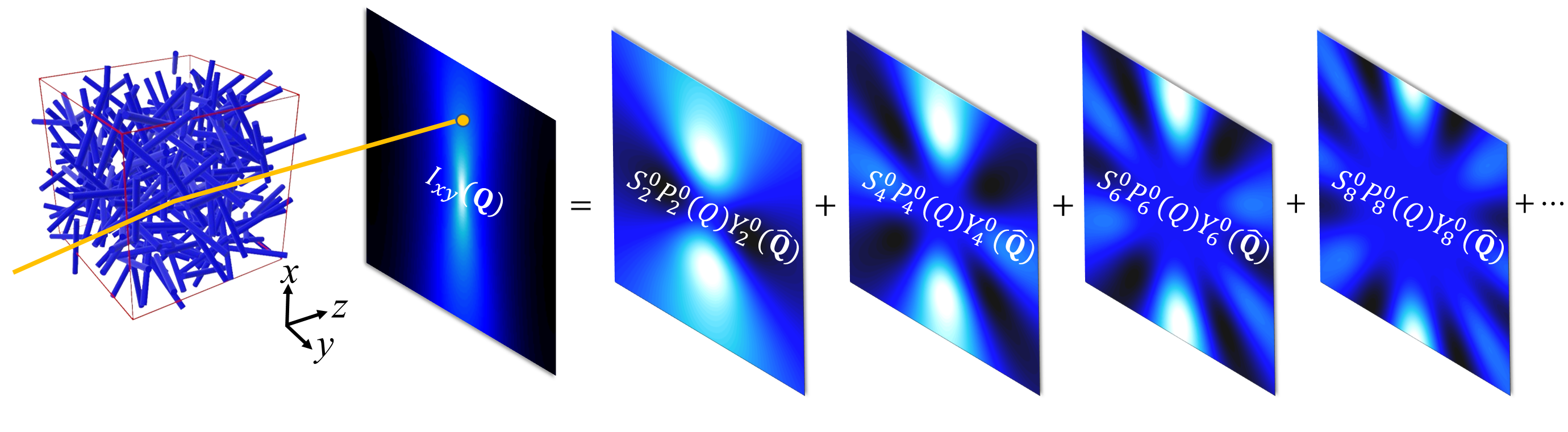}
}  
\caption{A demonstration of spectral decomposition as described in Eqn.~\eqref{eq:3.i2}. $I_{xy}(\textbf{Q})$ in Fig.~\ref{fig:2}(d) is used in this example.
}  
\label{fig:3}
\end{figure}

From the orthogonality of $Y_l^0$, $S_l^0$ required for reconstructing $f(\Omega)$ can be determined either from $I_{xy}(\textbf{Q})$ or $I_{xz}(\textbf{Q})$. For example, 
\begin{equation}
S_l^0 = \frac{1}{P_l^0(Q)}\int_{0}^{\frac{\pi}{2}}  d\theta \sin \theta I_{xy}(\textbf{Q})Y_l^0(\cos \theta).
\label{eq:3.i5}
\end{equation}
$P_l^0(Q)$ is required for extracting the ODF via Eqn.~\eqref{eq:3.b}. As a demonstration, we exemplify this with a rod, and the intra-rod spatial correlation function can be shown to take the following expression
\begin{equation}
P(\textbf{Q}, \Omega) = \left[2j_0 \left(\frac{QL_R}{2} \cos \beta \right)\frac{J_1(QR\sin{\beta})}{QR\sin{\beta}} \right]^2.
\label{eq:3.i6}
\end{equation}
Here, $j_0$ represents the spherical Bessel function of order $0$, $J_1$ is the first kind of Bessel function of order $1$, and $\beta$ denotes the angle between the scattering vector $\textbf{Q}$ and the principal axis of the tagged rod particle. $L_R$ and $R$ respectively denote the length and cross sectional radius of the rod, which can be determined from the rod suspensions in their quiescent state using scattering.

In conclusion, this subsection outlines the spectral decomposition described in Eqn.~\eqref{eq:3.i2}, as illustrated in Fig.~\ref{fig:3}.

\subsubsection{Spectra without Axial Symmetry}

Recent SANS measurements have unveiled a notable absence of axial symmetry in spectra derived from the flow-velocity gradient plane, exemplified by the structural analysis of worm-like micellar solutions within a Couette shear cell \cite{Wanger2014SM}. This absence poses a challenge for extracting the ODF. Unlike scenarios characterized by axial symmetry, where the ODF reconstruction relies only on spherical harmonic functions with zero degrees, the lack of mirror symmetry requires the inclusion of spherical harmonic basis functions with non-zero degrees ($m$) for ODF extraction. 

\begin{figure}
\centerline{
  \includegraphics[width=\linewidth]{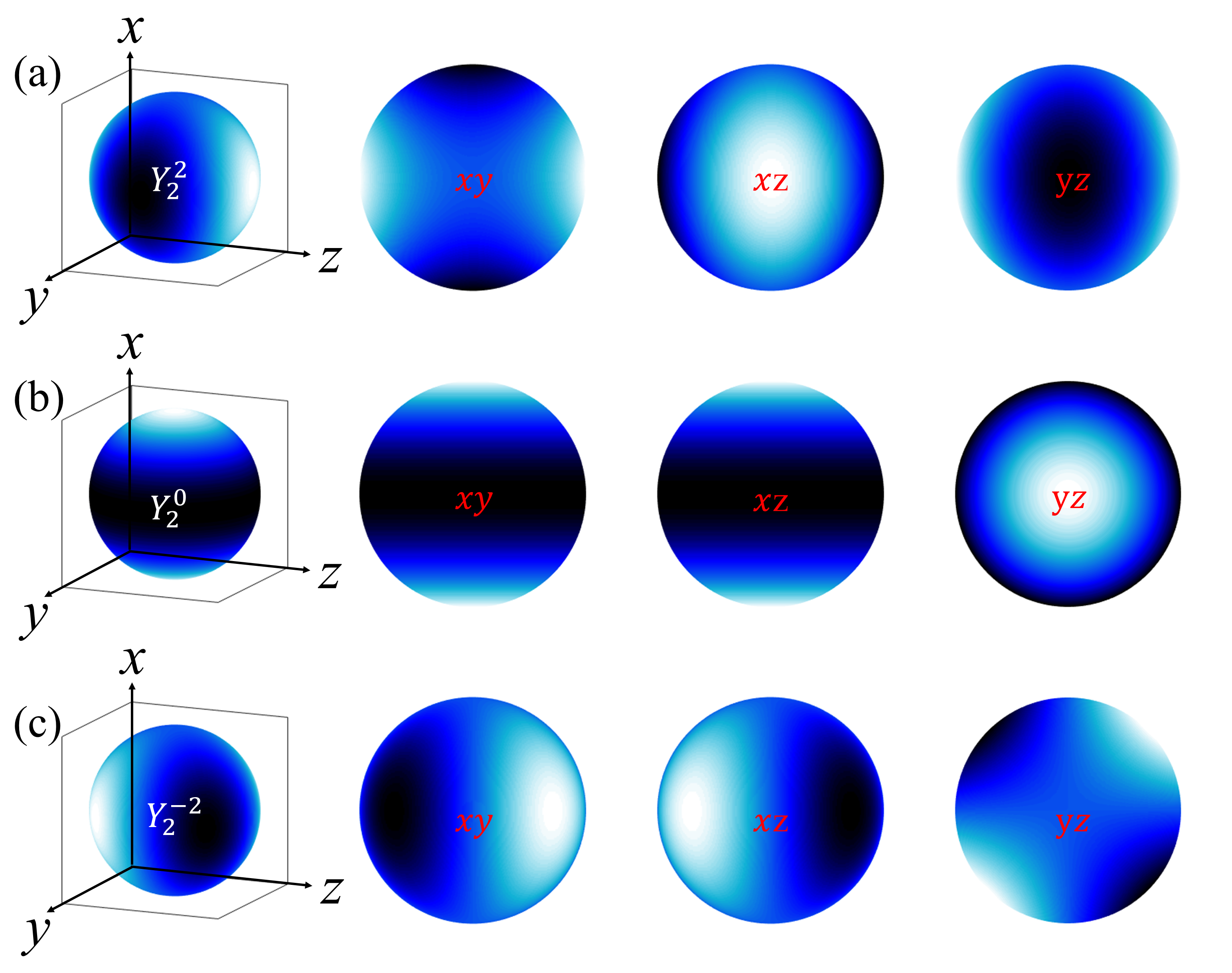}
}  
\caption{Visual representations of the first three real spherical harmonics, $Y_2^2$ (a), $Y_2^0$ (b), and $Y_2^{-2}$ (c), shown in Cartesian coordinates along with their projections on the $xy$, $xz$, and $yz$ planes.
}  
\label{fig:4}
\end{figure}

However, mathematical evidence indicates that these non-zero degree basis functions lose orthogonality on two-dimensional planes and exhibit coupling instead \cite{Huang2017}. To illustrate this phenomenon, Fig.~\ref{fig:4} presents visual representations of three second-degree real spherical harmonics, \(Y_2^2\) (a), \(Y_2^0\) (b), and \(Y_2^{-2}\) (c), in Cartesian coordinates. Their projections on the \(xy\), \(xz\), and \(yz\) planes are shown in the three entries of the same row, respectively. The loss of orthogonality of various spherical harmonic functions on two-dimensional planes can be intuitively appreciated by visual inspection. For instance, the orthogonality condition described by Eqn.~\eqref{eq:3.a1} no longer holds for \(Y_2^0\) and \(Y_0^0\) on the \(yz\) plane. Since the projection of \(Y_2^0\) is non-uniform but angularly isotropic, and \(Y_0^0\) is uniform, the integration in Eqn.~\eqref{eq:3.a1} will not be zero.  

The linear dependence of second-degree real spherical harmonics on two-dimensional planes, as illustrated in Fig.~\ref{fig:4}, can be further expressed mathematically. For instance, the relationships between $Y_0^0$, $Y_2^0$, and $Y_2^2$ on the $xy$-, $xz$-, and $yz$-planes are respectively given by:
\begin{eqnarray}
Y_2^0 &=& \sqrt{5}Y_0^0-\sqrt{3}Y_2^2, \: xy \mbox{ plane} \nonumber\\
Y_2^0 &=& \sqrt{5}Y_0^0+\sqrt{3}Y_2^2, \: xz \mbox{ plane} \nonumber\\
Y_2^0 &=& -\frac{\sqrt{5}}{2}Y_0^0. \: yz \mbox{ plane}
\end{eqnarray}
Consequently, in experimental studies governed by such complexities, the ODF cannot be uniquely determined from the collected scattering spectra due to this coupling complication.

\begin{figure}
\centerline{
  \includegraphics[width=\linewidth]{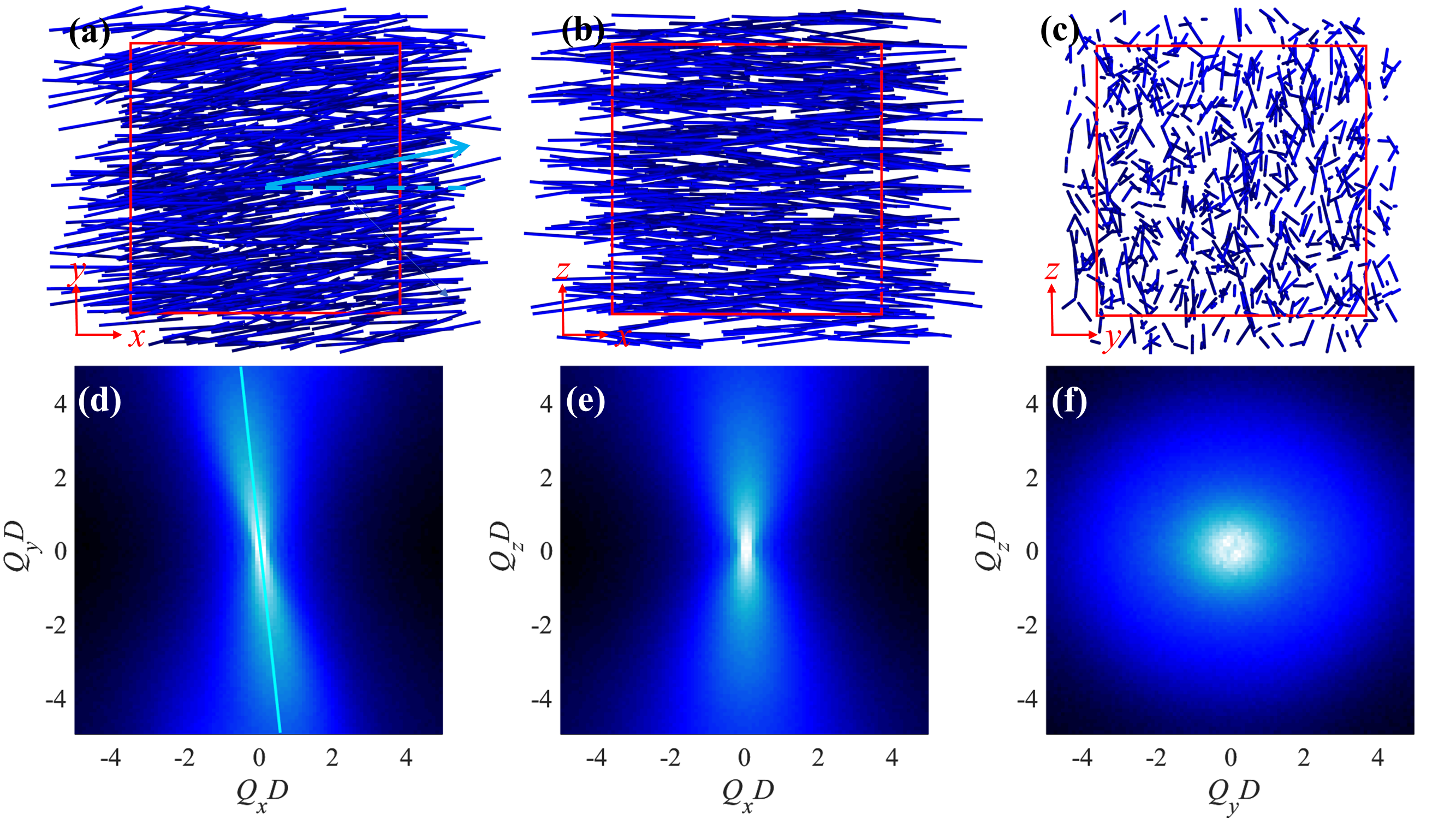}
}  
\caption{Snapshot of a sheared semi-dilute solution containing rigid rods with diameter $D$ and length $25D$. Real-space projections from the flow-velocity gradient ($x-y$), flow-vorticity ($x-z$), and velocity gradient-vorticity ($y-z$) planes are depicted in (a-c), respectively. Corresponding scattering spectra for these projections are shown in (d-f). The blue arrow in Panel (a) indicates the average tilt angle of flowing rods in real space. Consequently, the mirror symmetry of $I_{xy}(\textbf{Q})$ with respect to the axes of $Q_xD = 0$ and $Q_yD = 0$ is no longer preserved as shown in Fig.~\ref{fig:2}(d). 
}  
\label{fig:5}
\end{figure}

Having identified the issue as the coupling of $Y_l^m(\hat{\textbf{Q}})$ with non-zero $m$ on two-dimensional planes, one logical approach to mitigate this mathematical complexity is to conduct spectral analysis using an orthogonal set of spherical harmonic functions $Y_l^0$. To achieve this, on the pertinent flow plane—in our example, the $xy$ plane—a new principal axis, with respect to which $I_{xy}(\textbf{Q})$ demonstrates mirror symmetry, needs to be determined \cite{Huang2021}. 

From a mathematical standpoint, this process entails identifying a new reference frame aimed at eliminating the azimuthal angle $\phi$ dependence of $I_{xy}(\textbf{Q})$ in the lab frame. This new reference frame, represented by the light blue line in Fig.~\ref{fig:5}(d), aligns with the orientational direction of flowing particles (depicted as the $z'$ axis in Fig.~\ref{fig:1}), thereby introducing the desired axial symmetry. In this reference frame, the observed $\phi$ dependence becomes irrelevant, rendering only the terms with $m=0$ pertinent for spectral analysis. Consequently, the ODF relies solely on a new polar angle, $\Theta$, thereby bypassing the challenge of loss of orthogonality. This entails re-expressing Eqn.~\eqref{eq:3.d} as:
\begin{equation}
I(\textbf{Q}) = \sum_{l=0, 2, 4,\cdots}^{\infty} \mathbb{S}_l^0 P_l^0(Q) Y_l^0(\cos \Theta).
\label{eq:3.j}
\end{equation}
In this new reference frame, $\mathbb{S}_l^0$ denotes the order parameter. $Y_l^0(\cos \Theta)$ represents spherical harmonic functions of degree $l$ and order $0$ with the argument $\cos \Theta$. The ODF in this new reference frame takes the following expression: 
\begin{equation}
f(\Theta) = \sum_{l=0,2,4,\cdots}^{\infty} \mathbb{S}_l^0Y_l^0(\cos \Theta).
\label{eq:3.k}
\end{equation}
The orthogonality of $Y_l^0(\cos \Theta)$ requires that
\begin{equation}
\int_0^{2\pi} d \Phi \int_0^{\pi} d \Theta \sin \Theta Y_{l_1}^0(\cos \Theta) Y_{l_2}^0(\cos \Theta) = 4\pi\delta_{l_1,l_2},
\label{eq:3.l}
\end{equation}
where $l_1$ and $l_2$ are integers ranging from 0 to infinity. $\Phi$ is the azimuthal angle in the new reference frame.

This can be achieved by rotating the $x$-axis of the original reference frame, defined by the black axes, to align with the $x'$-axis of the new frame specified by the red axes, through an angle $\theta_t$, as illustrated in Fig. \ref{fig:1}. Axial symmetry is maintained within the $x'y'$ plane of the new reference frame, where the $x'$-axis aligns with the major axis of the elongated particle. 

From $\mathbb{R}$ given in Eqn.~\eqref{eq:2.0}, the relationships between the polar and azimuthal angles in the original reference frame and the new reference frame are determined to be:
\begin{eqnarray}
\cos \Theta &=& \cos \theta_t \cos \theta + \sin \theta_t \sin \theta \cos \phi \nonumber\\
\sin \Theta \cos \Phi &=& -\sin \theta_t \cos \theta + \sin \theta \cos \phi \cos \theta_t \nonumber\\
\sin \Theta \sin \Phi &=& \sin \theta \sin \phi
\label{eq:3.m}
\end{eqnarray}
Evidently, the quantity of interest for reconstructing the ODF is $\mathbb{S}_l^0$. Hence, the immediate question is: how can one derive $\mathbb{S}_l^0$ from the diverse projections of $I(\textbf{Q})$? 

From Eqns.~\eqref{eq:3.j}, \eqref{eq:3.k}, \eqref{eq:3.l}, and \eqref{eq:3.m}, $I_{xy}(\textbf{Q})$, $I_{xz}(\textbf{Q})$ and $I_{yz}(\textbf{Q})$ can be expressed as:
\begin{equation}
I_{xy}(\textbf{Q}) = \sum_{l=0,2,4,\cdots}^{\infty}\mathbb{S}_l^0P_l^0(Q)Y_l^0[\cos(\Theta \pm \theta_t)],
\label{eq:3.n}
\end{equation}
\begin{equation}
I_{xz}(\textbf{Q}) = \sum_{l=0,2,4,\cdots}^{\infty}\mathbb{S}_l^0P_l^0(Q)Y_l^0(\cos \theta_t \cos \Theta),
\label{eq:3.o}
\end{equation}
\begin{equation}
I_{yz}(\textbf{Q}) = \sum_{l=0,2,4,\cdots}^{\infty}\mathbb{S}_l^0P_l^0(Q)Y_l^0(\sin \theta_t \cos \phi),
\label{eq:3.p}
\end{equation}
In Eqns.~\eqref{eq:3.o} and \eqref{eq:3.p}, the arguments of $Y_l^0$ are not cosine functions, thus preventing the application of the orthogonality property provided in Eqn.~\eqref{eq:3.l} to derive $\mathbb{S}_l^0$ from either $I_{xz}(\textbf{Q})$ or $I_{yz}(\textbf{Q})$. However, through a variable transformation, the formalism can be adjusted to represent $Y_l^0$ with cosine functions of a single variable as its argument in Eqns.~\eqref{eq:3.n}. Subsequently, $\mathbb{S}_l^0$ can be obtained from $I_{xy}(\textbf{Q})$ by the following operation:
\begin{equation}
\mathbb{S}_l^0 = \frac{1}{P_l^0(Q)}\int_{\theta_t}^{\theta_t + \frac{\pi}{2}}  d\theta \sin \theta I_{xy}(\textbf{Q})Y_l^0(\cos \theta),
\label{eq:3.q}
\end{equation}
By substituting Eqn.~\eqref{eq:3.q} into Eqn.~\eqref{eq:3.k} and utilizing the expression of $P_l^0(Q)$ given in Eqn.~\eqref{eq:3.e}, the ODF can be reconstructed.

Before concluding this subsection, it is noteworthy to mention that the measurements of $I_{xy}(\textbf{Q})$ have long been acknowledged as pivotal for elucidating the connection with structural distortion and shear viscosity \cite{Solomon2005}. The calculation further underscores this assertion by showcasing the significance of the compatibility between the deformation gradient and the probed projection of $I(\textbf{Q})$ for the structural investigation of flowing materials. A Couette geometry shear cell has been designed and used for studying the critical flow-velocity gradient plane with SANS \cite{PorcarJoVE}.   

\subsubsection{Reconstructing the ODF Using the Principle of Maximum Probabilistic Entropy}

In practical spectral analysis, using an infinite number of spherical harmonic functions to reconstruct the ODF becomes impractical due to computational constraints and uncertainties associated with the measurement. This raises the question: how can $f(\Omega)$ be precisely inverted from scattering spectra while minimizing computational complexity? In response, one can construct the most probable ODF using the maximum probabilistic entropy formalism \cite{Kardar}. Below, we outline this mathematical strategy.

First the information entropy can be defined as 
\begin{equation}
-\langle \ln f(\Omega)\rangle = -\int d \Omega f(\Omega) \ln f(\Omega),
\label{eq:3.r}
\end{equation}
where $\langle \cdots \rangle$ denotes the average with respect to $f(\Omega)$. Using the method of Lagrange multipliers, we maximize the objective function,
\begin{equation}
H = -\langle \ln f(\Omega)\rangle + \sum_{l,m} \lambda_l^m[\langle Y_l^m (\Omega) \rangle - S_l^m],
\label{eq:3.s}
\end{equation}
where the constant $\lambda_l^m$ represents the Lagrange multiplier associated with the constraint $S_l^m$. Maximizing $H$ involves performing functional derivatives on  $H$ with respect to $f$ and setting them to zero. Namely, 
\begin{equation}
\delta H = 
 \int \mathrm{d}\Omega \delta f(\Omega) [-1 - \ln f(\Omega) + \lambda_l^m Y_l^m(\Omega)] = 0.
\label{eq:3.t}
\end{equation}
This equality holds only when the quantity inside the bracket $\langle \cdots \rangle$ equals zero and $f(\Omega)$ satisfies the constraint equation of
\begin{equation}
S_l^m = \frac{1}{4\pi} \int d \Omega f(\Omega) Y_l^m(\Omega).
\label{eq:3.u}
\end{equation}

The resulting ODF is thus demonstrated to assume the following expression:
\begin{equation}
f(\Omega) = \exp\left[\sum_{l = 0}^{\infty} \sum_{m = -l}^l \lambda_l^m Y_l^m(\Omega)\right].
\label{eq:3.v}
\end{equation}
In the reference frame where ODF is axially symmetrical along the selected principal direction, Eqn.~\eqref{eq:3.v} can further be simplified as
\begin{equation}
f(\Theta) = \exp\left[\sum_{l = 0}^{\infty} \Lambda_l^0 Y_l^0(\Theta)\right].
\label{eq:3.w}
\end{equation}

The validity of this approach is verified through numerical testing on various flow systems characterized by different ODFs with different analytical expressions, such as uniaxial extension homogeneous flow \cite{Doi1986, Rubinstein2003, Hakansson2016}, Doi–Edwards ODF \cite{Doi1986}, and Kramer’s shear flow \cite{Doi1986, Rubinstein2003}. The reconstructed ODFs based on $\Lambda_0^0$ and $\Lambda_2^0$ \cite{Huang2021} demonstrate quantitative agreement with the corresponding target $f(\Omega)$ across different analytical expressions \cite{Huang2021}. 

\begin{figure}
\centerline{
  \includegraphics[width=\linewidth]{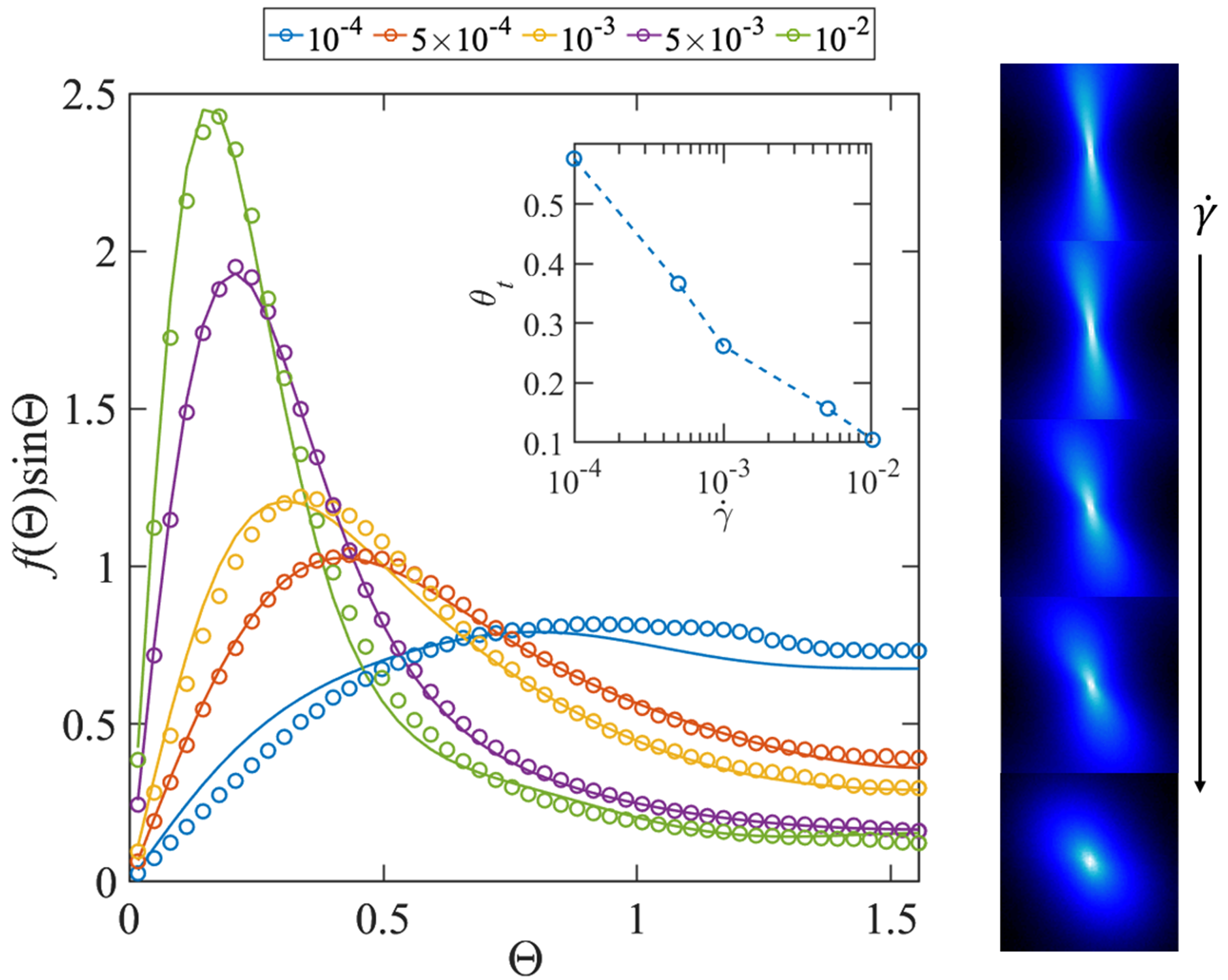}
}  
\caption{The orientational distribution functions, multiplied by $\sin{\Theta}$ in the axially symmetrical frame, were obtained from dissipative particle dynamics simulations of 216 rigid rods in a semi-dilute solution under shear at five different shear rates. The results from Eqn.~\eqref{eq:3.w} are represented by the solid lines. Given that the probability of finding rods at a specific $\Theta$ is expressed as $\mathrm{d}\Omega f(\Omega) \equiv \mathrm{d}\Theta \sin{\Theta} f(\Theta)$, $f(\Theta)\sin{\Theta}$ is used as the vertical axis. The right panels display the corresponding $I_{xy}(\textbf{Q})$. The inset shows the tilt angles $\theta_t$, represented by the dashed line, as a function of shear rate. The unit of $\Theta$ is radians.
}  
\label{fig:6}
\end{figure}

Furthermore, computational benchmarking using DPD simulations was conducted to assess the numerical accuracy of the inverted ODF of sheared rigid rods, whose mathematical expression is not analytically known. As illustrated in Fig.~\ref{fig:6}, there is a quantitative agreement between the inverted $f(\Omega)$, derived from spectral analysis according to Eqn.~\eqref{eq:3.w}, and the $f(\Omega)$ obtained directly from trajectory analysis within the examined range of shear rates. This agreement highlights the validity and mathematical effectiveness of the maximum probabilistic entropy formalism in facilitating the inversion of $f(\Omega)$ for general flowing systems using small angle scattering techniques.

\subsubsection{Scalar Descriptor Approach for Quantifying Particle Alignment}

In the sections above, we have conducted a thorough investigation into both parametric and non-parametric methods for deducing the ODF from the scattering patterns of flowing objects. Here we examine a widely adopted spectral analysis technique known as the order parameter, which originated from methods designed to assess axial symmetry in nematic liquid crystal phases \cite{Leadbetter1979, Leadbetter1984}. Theoretical studies on anisotropic lyotropic phases further highlight the use of order parameters in assessing system free energy and predicting phase transitions, highlighting its importance in experimental studies.

The order parameter approach shares close mathematical connections with the previously discussed non-parametric method, as illustrated by Eqn. \eqref{eq:3.i}. In theoretical analyses of flowing elongated particles, the relevant order parameter is $S_l^0$, not $\hat{S}_l^0(Q)$ in reciprocal space. In diffraction experiments involving materials with smectic-like local ordering \cite{Roe, Leadbetter1979, Leadbetter1984, Caspar1988, Deutsch1991, Caspar1993, vanGurp, Mochrie2003, Heiney2006, Agra-Kooijman2018, Sims2019}, the intensity distribution occurs within a narrow range of $Q$. In such cases, the radial component of intra-particle correlation, $P_0^0 (Q)$ and $P_l^0 (Q)$ in Eqn.~\eqref{eq:3.i}, can be effectively approximated as a delta function. Consequently, the influence of the form factor decreases, and the dependence of $\hat{S}_l^0 (Q)$ on $Q$ becomes insignificant, indicating $\hat{S}_l^0(Q) \sim S_l^0$. 

However, in general small angle scattering experiments, the contribution of intra-particle spatial correlation to the measured scattering intensity cannot be disregarded. It has been recognized \cite{Winnik2011} that the failure to account for this intra-particle spatial correlation may lead to erroneously regarding $\hat{S}_l^0(Q)$, obtained directly from small angle scattering data analysis, as the pivotal quantity of $S_l^0$ \cite{Burghardt1996, Burghardt1998, Lindner1998, Hsiao2005, Hsiao2006, Porcar2008, Porcar2010, Ch2011, Hakansson2014, Haywood2017, Lang2016, Lang2019, Li2020, Munier2022, Tabor2022, Helgeson2024}. For an in-depth exploration of the mathematical relationships between $S_l^0$ and its corresponding $\hat{S}_l^0(Q)$, readers are directed to the works by Lovell \cite{Lovell} and Huang \cite{Huang2021}.

\begin{figure}
\centerline{
  \includegraphics[width=\linewidth]{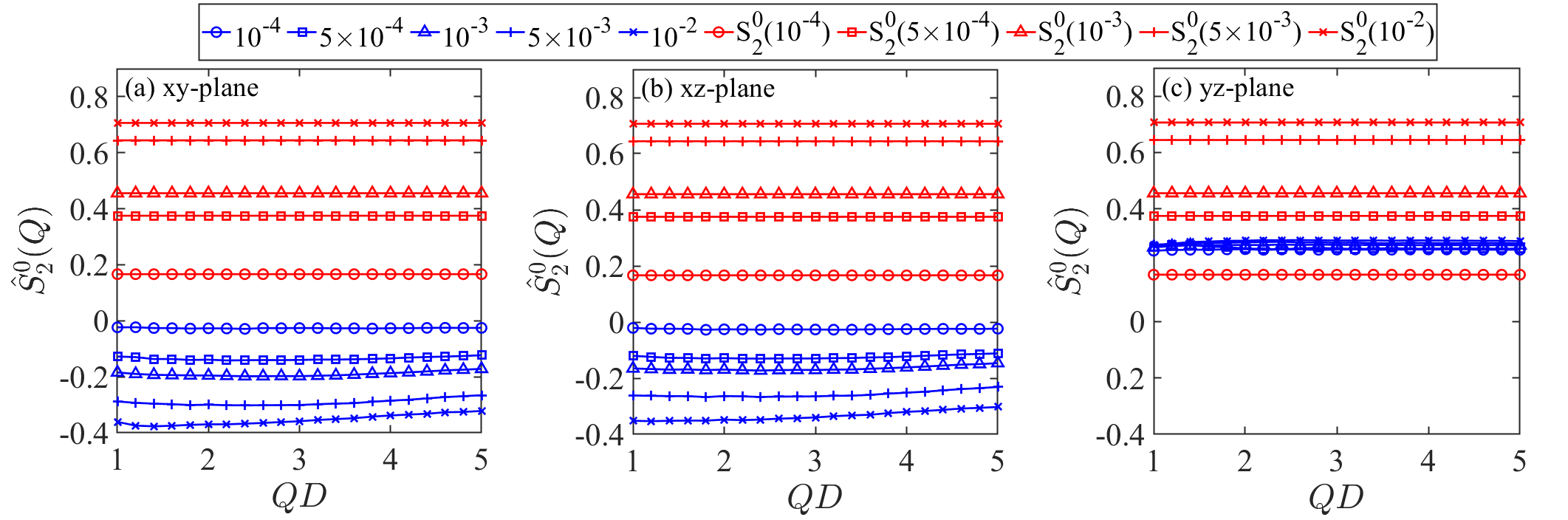}
}  
\caption{Scattering order parameters $\hat{S}_2^0(Q)$ (blue symbols) and their corresponding order parameters $S_2^0$ (red symbols) derived from the scattering intensities $I_{xy}(\textbf{Q})$ (a), $I_{xz}(\textbf{Q})$ (b), and $I_{yz}(\textbf{Q})$ (c) of semi-dilute rigid rod suspensions under varying shear rates, as discussed in Fig.~\ref{fig:6}. 
}  
\label{fig:7}
\end{figure}

To illustrate this point, Fig.~\ref{fig:7} presents \(\hat{S}_2^0(Q)\) (blue symbols) and the corresponding \(S_2^0\) (red symbols) derived from the scattering intensities \(I_{xy}(\mathbf{Q})\) (a), \(I_{xz}(\mathbf{Q})\) (b), and \(I_{yz}(\mathbf{Q})\) (c) of semi-dilute rigid rod suspensions under varying shear rates, as discussed in Fig. \ref{fig:6}. Due to the presence of intra-particle spatial correlation, the \(Q\) dependence of \(\hat{S}_2^0(Q)\) extracted from \(I_{xy}(\mathbf{Q})\) and \(I_{xz}(\mathbf{Q})\) is clearly discernible, particularly when the shear rate is higher than \(10^{-3}\). In this computational benchmarking, \(\hat{S}_2^0(Q)\) extracted from \(I_{yz}(\mathbf{Q})\) obtained from different shear rates are seen to nearly collapse around \(0.2\). Its \(Q\) dependence is not discernible. In a real scattering experiment, determining the absolute intensity of \(I_{yz}(\mathbf{Q})\) is also particularly challenging due to the Couette geometry of the shear cell. As a result, it is practically not feasible to extract the order parameter of sheared systems from the measurement of \(I_{yz}(\mathbf{Q})\).  
                                                     
It is worth noting that it has been suggested \cite{Winnik2011} that 
\begin{equation}
\lim_{Q\to\infty} \hat{S}_2^0(Q) = - \frac{1}{2} S_2^0.
\label{eq:3.w1}
\end{equation}

\begin{figure}
\centerline{
  \includegraphics[width=\linewidth]{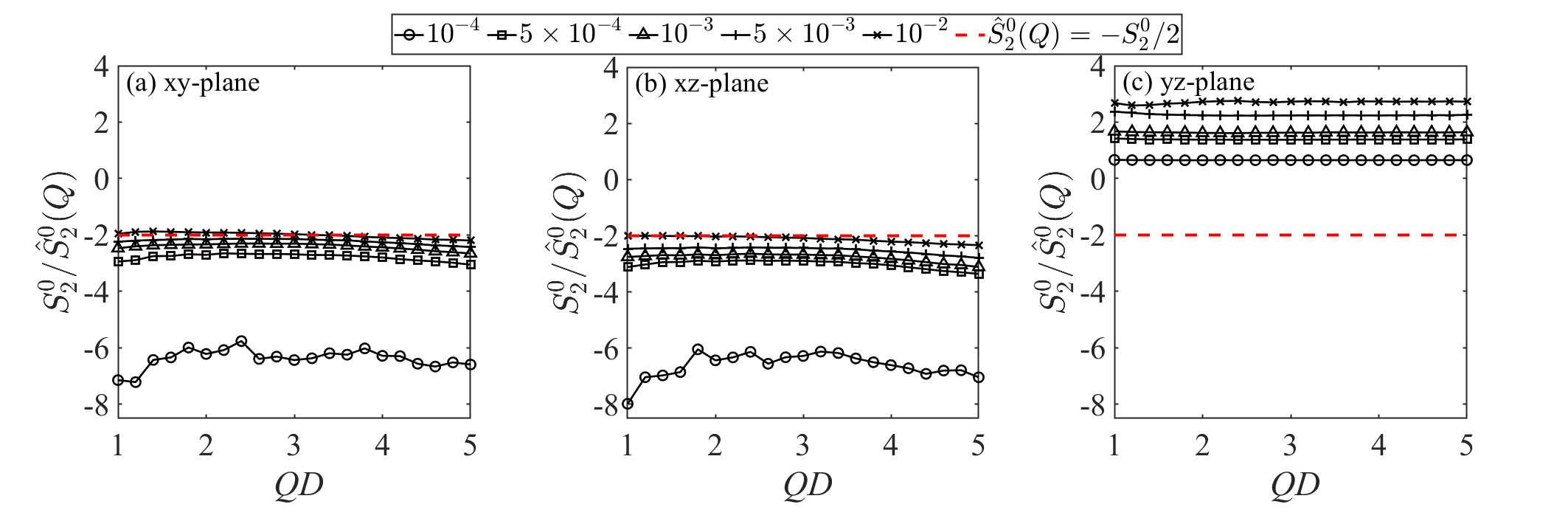}
}  
\caption{The black symbols represent the values of $\frac{S_2^0}{\hat{S}_2^0(Q)}$ derived from the data presented in Fig.~\ref{fig:7}. The red dashed lines indicate the asymptotic value of $-2$ in the high $Q$ limit suggested by Gilroy and colleagues \cite{Winnik2011}. In panel (c), the inadequacy of using $I_{yz}(\textbf{Q})$ to extract the order parameter of sheared systems is once again highlighted.
}  
\label{fig:8}
\end{figure}

Fig.~\ref{fig:8} illustrates the values of $\frac{S_2^0}{\hat{S}_2^0(Q)}$ derived from the data presented in Fig.~\ref{fig:7}. Particularly in the higher \(Q\) range, the subtraction of incoherent scattering ($I_{inc}$) from the measured scattering intensity inevitably leads to increased statistical uncertainties due to the dominant influence of $I_{inc}$ in this region. It becomes evident that the relationship defined in Eqn.~\eqref{eq:3.w1} does not strictly hold within the range of \(QD < 5\) in our computational benchmarking. Furthermore, results obtained from systems subjected to lower shear rates show greater deviation from the asymptotic value of $-2$, indicating that the ODF cannot be reliably reconstructed solely from \(\hat{S}_2^0(Q)\) using the formula provided in  Eqn.~\eqref{eq:3.w1}. This benchmarking underscores that the optimal method for extracting the order parameter from the anisotropy of the measured spectra should rely on  Eqn.~\eqref{eq:3.i}.

Another commonly used scalar description of particle orientation is the alignment factor \cite{Wagner1996, Wagner1997, Wagner2005, Liberatore2006, Wagner2006, Wagner2009, Wagner2009_2, Liberatore2009, Wagner2016, Helgeson2016, Porcar2022, Helgeson2024} which takes the following expression:
\begin{equation}
A_f(Q) = \frac{\int_0^{2 \pi} d \phi I_p(\textbf{Q}) \cos (2\phi)}{\int_0^{2 \pi} d \phi I_p(\textbf{Q})}.
\label{eq:3.x}
\end{equation}
In this context, $I_p(\textbf{Q})$ represents the projection of the three-dimensional $I(\textbf{Q})$ onto a specific flow plane. In its original definition, $\phi$ signifies the azimuthal angle relative to the $x$-axis. However, the efficacy of the direct spectral weighting manipulation detailed in Eqn.~\eqref{eq:3.x} in offering quantitative insights into the ODF is called into question by the $Q$ dependence displayed by $A_f(Q)$, given that the ODF of rigid elongated objects inherently remains independent of $Q$.    

To support this statement, we scrutinize the mathematical characteristics of $A_f(Q)$ from the perspective of basis expansion: Considering the case, where $I_p(\textbf{Q})$ is axially symmetric to $x$-axis, the integrand of $A_f (Q)$ can be reduced to $[0,\pi]$. On the $xy$-plane, $\phi$ is the angle between $\textbf{Q}$ and the flow. Therefore, $I_p(\textbf{Q}) = I_{xy}(\textbf{Q})$ can be expressed as
\begin{equation}
I_{xy}(\textbf{Q}) = \sum_{l=0,2,4,\cdots}^{\infty} I_l^0(Q)Y_l^0(\cos{\phi}).
\label{eq:3.y}
\end{equation}
In Eqn.~\eqref{eq:3.y}, $l$ are even integers and $I_l^0(Q) = \frac{1}{2}\int_0^{\pi} d\phi \sin{\phi} I_{xy}(\textbf{Q})Y_l^0(\cos{\phi})$. From Eqns.~\eqref{eq:3.x} and ~\eqref{eq:3.y}, it is found that 
\begin{equation}
A_f(Q) = 1 - 2 \sum_{n=0}^{\infty} \binom{\frac{1}{2}}{n} \frac{\int_0^{\pi} \mathrm{d}\phi \sin \phi I_{xy}(\textbf{Q}) (-\cos^2\phi)^n}{\int_0^{\pi} d\phi I_{xy}(\textbf{Q})},
\label{eq:3.z}
\end{equation}
where $\binom{\frac{1}{2}}{n}$ denotes the binomial coefficient. The term $(- \cos ^2 \phi)^n$ within the integrand of Eqn.~\eqref{eq:3.z} encompasses the information pertaining to all $I_l^0 (Q)$ where $l \leq 2n$. For example, when $n=1$, $A_f(Q) = \frac{2}{15}\frac{5I_0^0(Q) - 2I_2^0(Q)}{\int_0^\pi d\phi I_{xy}(\textbf{Q})}$. This basic illustration suggests that determining $S_2^0$ from $A_f (Q)$ is mathematically challenging without imposing a specific functional form of the ODF. The claim that $A_f(Q)$ is equivalent to $-S_2^0$ in the high $Q$ limit \cite{Wagner1996} may therefore need further consideration. 

This analysis indicates that $A_f(Q)$ provides qualitative insights into the alignment of rods rather than precise quantitative information. The primary reason is the absence of an established rule or method to determine the appropriate $Q$ value at which to sample the scattering intensity for calculating the scalar descriptor $A_f(Q)$. This lack of a consistent guideline makes it challenging to obtain precise quantitative information.

In terms of extracting the ODF from spectral anisotropy, relying solely on $A_f(Q)$ appears to be less effective. This observation aligns with the understanding that Eqn.~\eqref{eq:3.x} suggests limitations in $A_f(Q)$ for capturing three-dimensional spatial information, primarily due to its omission of the polar angle component in the formulation of the three-dimensional differential solid angle. The reason for this limitation is straightforward: Unlike the azimuthal angle, which is defined within a specific flow plane, the polar angle is defined relative to a specific axis and is not confined to any plane.

\subsubsection{Assessing Feasibility in Experimental Scattering Data Analysis}

To demonstrate the feasibility of our proposed inversion method, we conducted rheo-SANS measurements on a well-studied wormlike micellar system: Aqueous solutions of cetylpyridinium chloride (CPyCl) and sodium salicylate (NaSal) in a 2:1 molar ratio. Samples were prepared with a surfactant concentration of 6 \% by mass and a salt concentration of 0.5 M NaCl in D$_{2}$O. The SANS experiment was performed using the D22 large dynamic range small-angle diffractometer at the Institut Laue-Langevin (ILL). In this study, we used a wavelength of 6 Å and a sample to detector distance of 11.2 m with a 11.2 m collimation (source to sample distance). The probed $Q$ range was from 0.001 \r{A}$^{-1}$ to 0.06 \r{A}$^{-1}$. A Couette geometry shear cell \cite{PorcarJoVE} was employed to probe the structural information projected onto the flow-velocity gradient ($xy$) plane.

\begin{figure}
\centerline{
  \includegraphics[width=\linewidth]{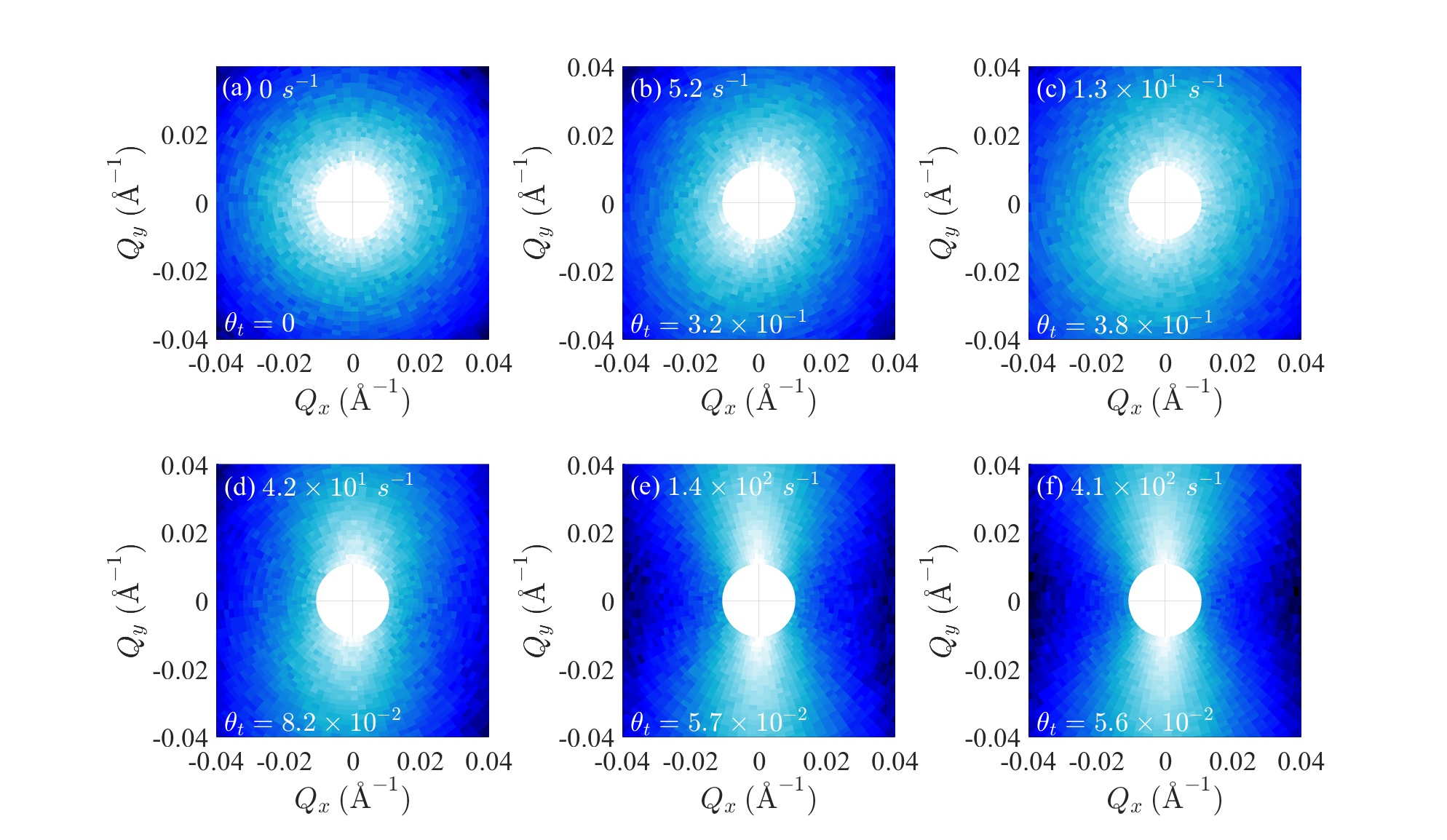}
}  
\caption{$I_{xy}(\mathbf{Q})$ of the sheared CpyCl/NaSal solution at five different shear rates $\dot{\gamma}$: 5.2 s$^{-1}$ (b), 13 s$^{-1}$ (c), 42 s$^{-1}$ (d), 140 s$^{-1}$ (e), and 410 s$^{-1}$ (f), along with the result obtained from the quiescent state (a). The tilt angle $\theta_t$ is also presented.
}  
\label{fig:9}
\end{figure}

Fig.~\ref{fig:9}(b)-(f) presents the $I_{xy}(\mathbf{Q})$ of the sheared CpyCl/NaSal solution at five distinct shear rates: 5.2, 13, 42, 140, and 410 s$^{-1}$. Panel (a) displays the result from the quiescent state for comparison. As the shear rate increases, $I_{xy}(\mathbf{Q})$ shows a more pronounced anisotropy. The loss of axial symmetry in the scattering spectra is quantitatively characterized by the evolution of $\theta_t$. Specifically, $\theta_t$ decreases from $3.2 \times 10^{-1}$ radians at a shear rate of 5.2 s$^{-1}$ to $5.6 \times 10^{-2}$ radians at a shear rate of 410 s$^{-1}$. This trend underscores the significant impact of shear rate on the structural organization within the solution.  

\begin{figure}
\centerline{
  \includegraphics[width=\linewidth]{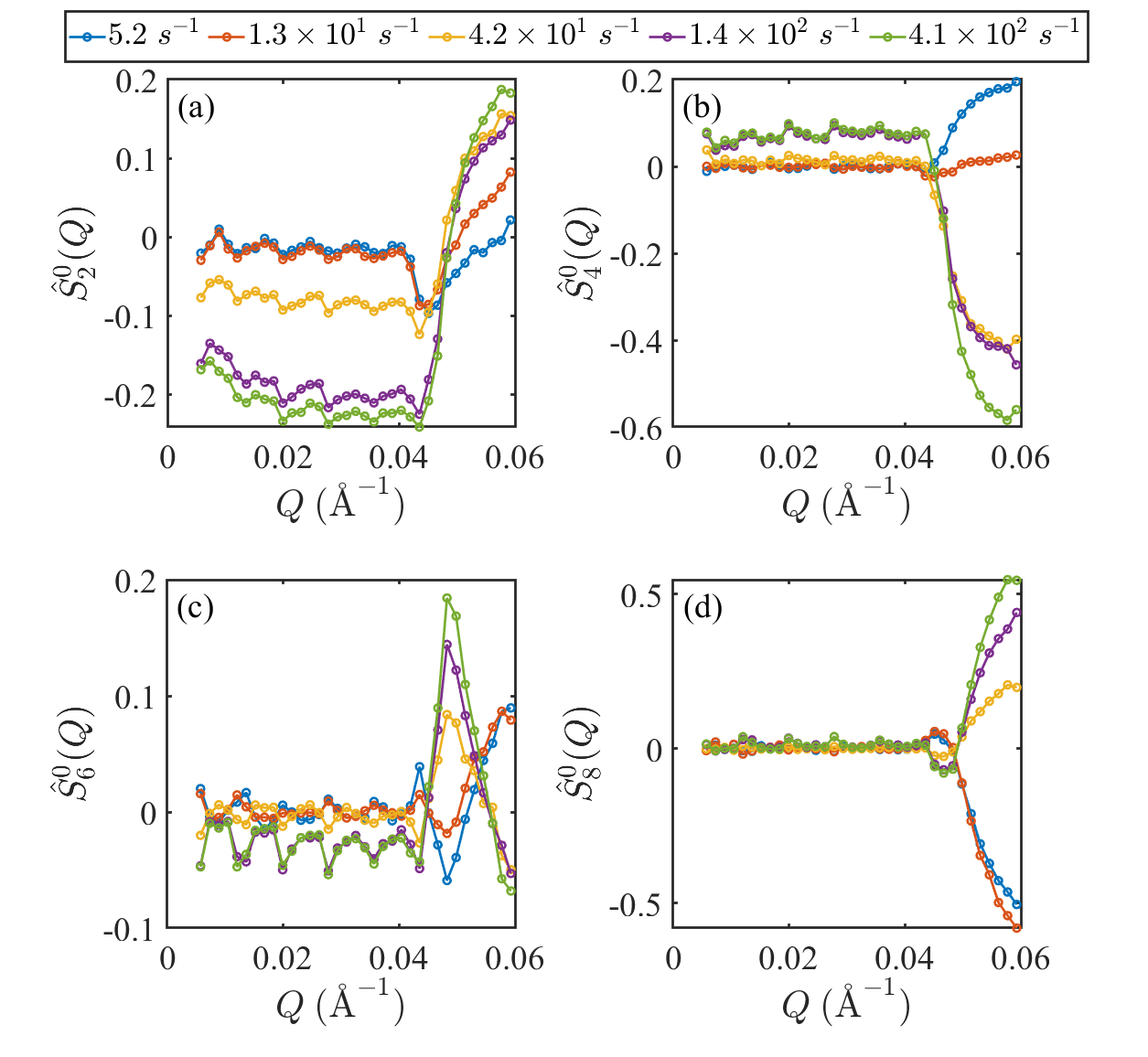}
}  
\caption{Scattering order parameters $\hat{S}_2^0(Q)$ (a), $\hat{S}_4^0(Q)$ (b), $\hat{S}_6^0(Q)$ (c), and $\hat{S}_8^0(Q)$ (d) extracted from the data in Fig.~\ref{fig:9} using Eqn.~\eqref{eq:3.g}. 
}  
\label{fig:10}
\end{figure}

The extracted $\hat{S}_2^0(Q)$, $\hat{S}_4^0(Q)$, $\hat{S}_6^0(Q)$, and $\hat{S}_8^0(Q)$ from Fig.~\ref{fig:9} via Eqn.~\eqref{eq:3.g} are presented in Fig.~\ref{fig:10}. Only the scattering order parameters within the $Q$ range of 0.01 \r{A}$^{-1}$ $< Q <$ 0.04 \r{A}$^{-1}$ should be used for discussion, as the detector edge effect becomes significant when $Q >$ 0.045 \r{A}$^{-1}$. Judging by the amplitude of the extracted scattering order parameters, it is clear that only $\hat{S}_2^0(Q)$, $\hat{S}_4^0(Q)$, and $\hat{S}_6^0(Q)$ are relevant for reconstructing the ODFs of sheared CPyCl/NaSal wormlike micelles.

\begin{figure}
\centerline{
  \includegraphics[width=\linewidth]{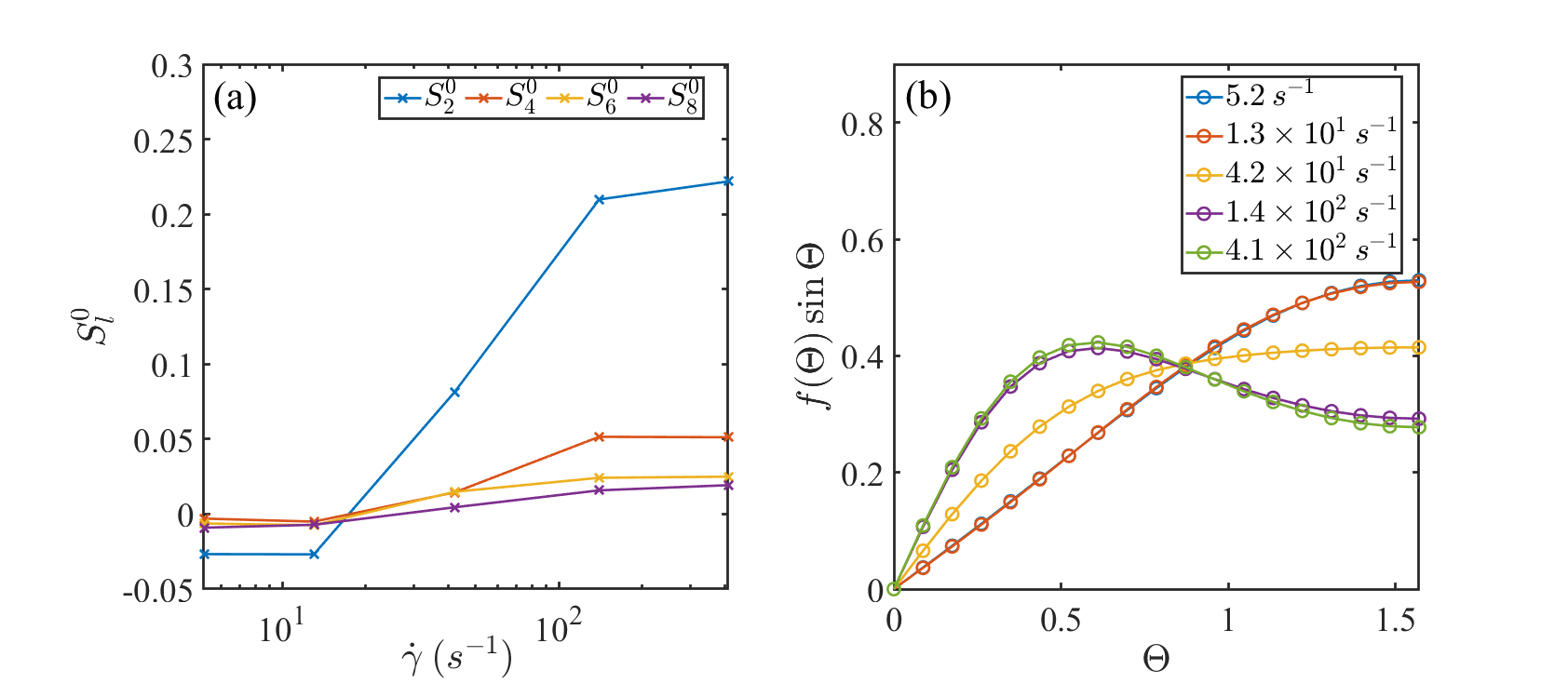}
}  
\caption{(a) Order parameters $S_l^0$ calculated from the corresponding scattering order parameters $\hat{S}_l^0(Q)$ presented in Fig.~\ref{fig:10} using Eqn.~\eqref{eq:3.i}. (b) Reconstructed ODFs at different shear rates using Eqn.~\eqref{eq:3.w}.
}  
\label{fig:11}
\end{figure}

Fig.~\ref{fig:11}(a) presents the $S_l^0$ calculated from the corresponding $\hat{S}_l^0(Q)$ using Eqn.~\eqref{eq:3.i}. In this calculation, the persistence length $b$ of CPyCl/NaSal wormlike micelles is 500 \r{A}, and the radius of the cross-section is 20 \r{A}. The dominant role of $S_2^0$ and the significant contribution of $S_4^0$ are clearly revealed. Fig.~\ref{fig:11}(b) displays the reconstructed ODFs using Eqn.~\eqref{eq:3.w} at different shear rates. Consistent with the characteristic development of $I_{xy}(\textbf{Q})$ displayed in Fig.~\ref{fig:9}, the progressively increasing shear-induced alignment is quantitatively indicated by the steady shift of the peak value of $f(\Theta)\sin \Theta$ from approximately 1.5 radians at a shear rate of 5.2 s$^{-1}$ to approximately 0.5 radians at a shear rate of 410 s$^{-1}$. 

This example demonstrates the efficacy of the nonparametric approach, leveraging the mathematical principles of basis expansion and maximum probabilistic entropy, to accurately extract the ODF of flowing elongated objects from their experimental SANS data, all while avoiding the constraints imposed by predefined models.

\subsubsection{Insights on Data Analysis from Ultra Small Angle Scattering (USAS) Studies of Flowing Materials}

The Bonse-Hart ultra small angle neutron and X-ray scattering instruments \cite{Bonse-Hart} have been used to carry out structural studies of objects that have dimensions larger than those available with conventional SAS measurements \cite{Burdette-Trofimov1, Burdette-Trofimov2, Burdette-Trofimov3, Burdette-Trofimov4, Burdette-Trofimov5}. The Q-range available with the Bonse-Hart instruments is useful for characterizing the alignment of rod-like materials under flow when the dimensions of rods exceed the size scale available in SAS. 

\begin{figure}
\centerline{
  \includegraphics[width=\linewidth]{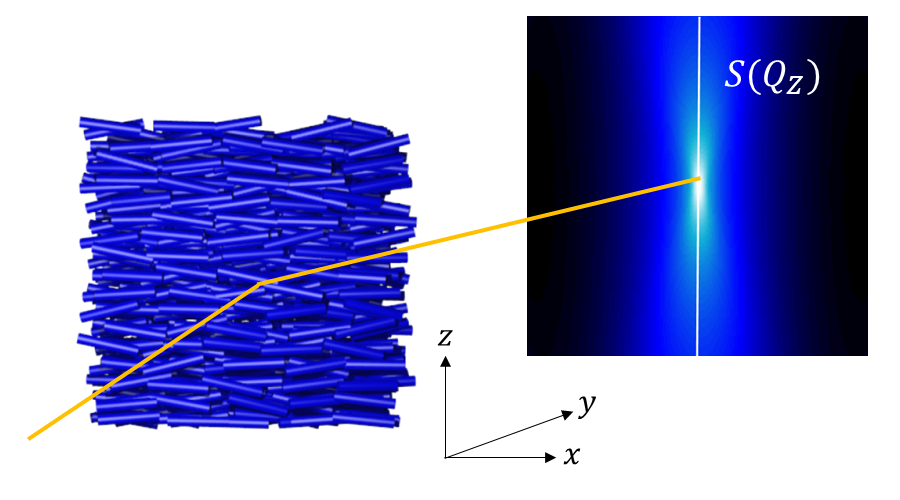}
}  
\caption{A schematic illustration depicting a USAS measurement of a rod-like suspension under steady shear. The slit geometry restricts structural information to the $xz$ plane along the $Q_z$ direction. Radiation is incident along the $y$ direction, with the slit's long axis oriented along the $x$ direction. The yellow line indicates the directions of incident and scattered radiation, with the incident radiation aligning with the $y$ axis.
}  
\label{fig:12}
\end{figure}

To accomplish the measurement of scattering at very low angles with a high angular resolution, the size and angular divergence of the incoming beams need to be restricted, leading to the loss of incoming beam flux. To balance the flux of incoming beams and the angular resolution necessary for the measurement at a USAS range, a slit geometry is often employed, particularly when the neutron is used as a probe. This geometry helps achieve good angular resolution in one direction but leads to a poor resolution in the other direction. Additionally, structural insights are limited to the $xz$ plane when a rheometer equipped with a Couette geometry shear cell is used to apply shear strain to the materials. It is crucial to determine how these factors restrict quantitative insights into the alignment of materials under steady shear.

Fig.~\ref{fig:12} shows a schematic illustration of USAS measurements, employing slit geometry, conducted on a rod-like suspension under steady shear. In the context of Eqn.~\eqref{eq:3.i3}, the measured scattering intensity of a solution containing rods with a length $L_R$ and a radius of cross section $R$ can be expressed as:  
\begin{eqnarray}
I_{xz}(Q_z) &=&
\int \mathrm{d}\Omega P(Q_z, \Omega) f(\Omega) \nonumber\\
&=& \int \mathrm{d}\Omega \left[2j_0 \left(\frac{Q_zL_R}{2} \cos \beta \right)\frac{J_1(Q_zR\sin{\beta})}{Q_zR\sin{\beta}} \right]^2 f(\Omega) \nonumber\\
&\sim& P_0^0(Q) + S_2^0P_2^0(Q).
\label{eq:3.usas1}
\end{eqnarray}
Eqn.~\eqref{eq:3.usas1} illustrates the capabilities and limitations of USAS with slit geometry for the quantitative investigation of the alignment of flowing rods: One can obtain a real-space order parameter $S_2^0$ when the geometric shape of the constituent particle is known in advance. In systems characterized by spectra exhibiting axial symmetry, as illustrated in Fig.~\ref{fig:2}, the ODF can be reconstructed using maximum probabilistic entropy principles, given the initial two real-space order parameters $S_0^0$ and $S_2^0$, as specified by Eqn.~\eqref{eq:3.w}.  

\begin{figure}
\centerline{
  \includegraphics[width=\linewidth]{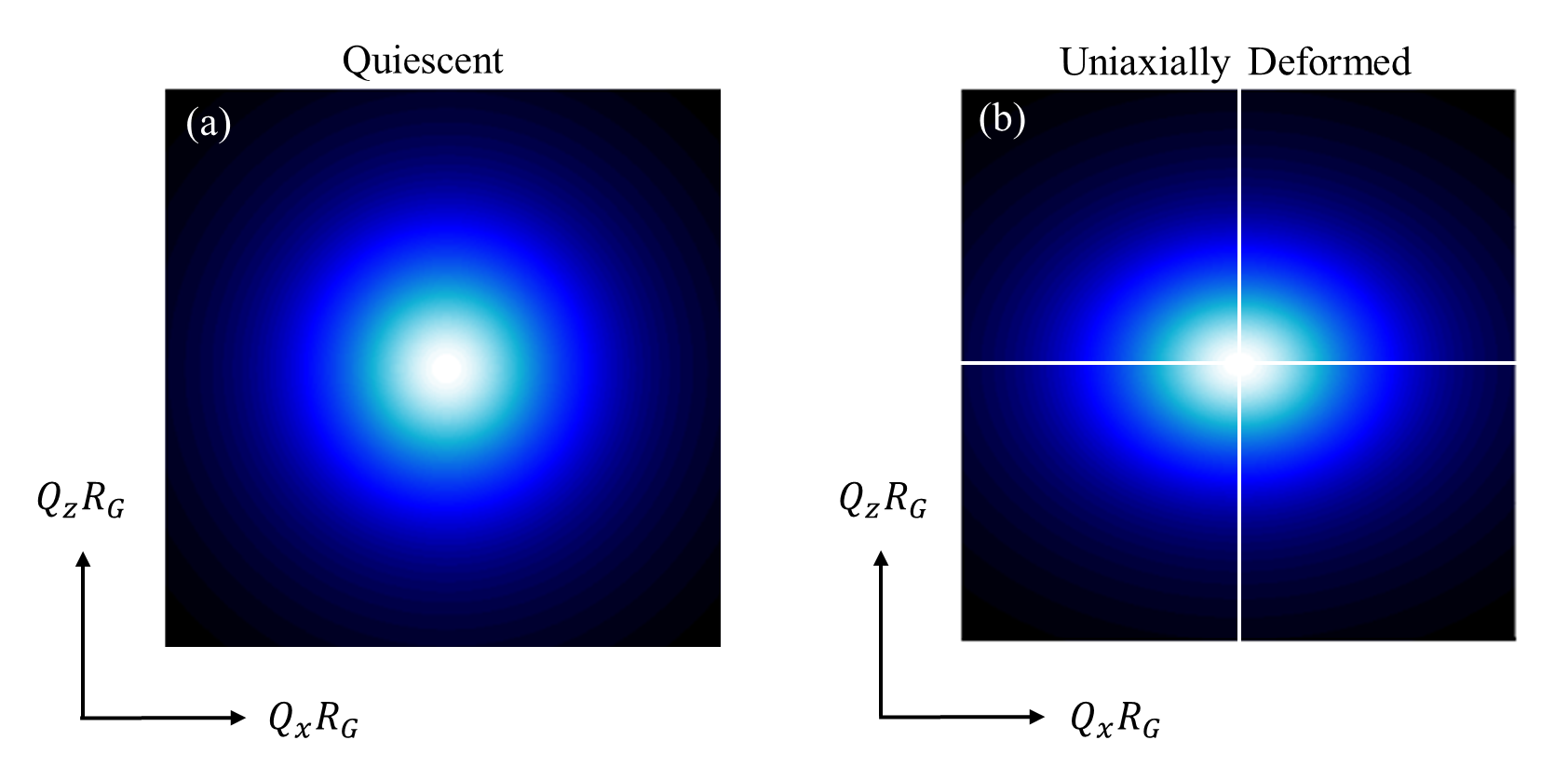}
}  
\caption{Scattering intensity of (left) a polymer melt at its quiescent state and (right) a stretched polymer melt. The stretching direction in real space aligns with the $z$ direction.
}  
\label{fig:13}
\end{figure}

Another intriguing avenue of research in USANS involves exploring the deformation characteristics of uniaxially stretched polymer melts, particularly those that are appropriately isotopically labeled. By rotating the sample, neutron scattering intensity both parallel and perpendicular to the direction of stretching can be obtained. Assuming the direction of stretching in real space aligns with the $z$ direction, via gyration tensor calculation \cite{HuangJPCL}, the scattering intensities along the $Q_z$ and $Q_x$ directions depicted in Fig.~\ref{fig:13} can be expressed as follows:   
\begin{equation}
I(Q_z) = 1 - Q^2(1 + \epsilon)^2R_G^2 + O (Q^4),
\label{eq:3.usas2}
\end{equation}
and
\begin{equation}
I(Q_x) = 1 - \frac{Q^2R_G^2}{1 + \epsilon} + O (Q^4).
\label{eq:3.usas3}
\end{equation}
where $R_G$ represents the radius of gyration at the quiescent state and $\epsilon$ denotes the engineering strain. For stretched polymers experiencing affine deformation, Eqns.~\eqref{eq:3.usas2} and ~\eqref{eq:3.usas3} give 
\begin{equation}
\frac{I(Q_z)-1}{I(Q_x)-1} \approx (1 + \epsilon)^3.
\label{eq:3.usas4}
\end{equation}
Eqn.~\eqref{eq:3.usas4} demonstrates how the strain can be conveniently extracted from the scattering intensities parallel and perpendicular to the direction of stretching. 

\section{Potential Future Research Directions}

In this section, we emphasize several research directions deemed promising for investigation. Drawing from the insights presented in the preceding section, it becomes evident that the potential of parametric regression analysis is somewhat limited due to the challenge of ascertaining the analytical expression of the ODF \textit{a priori}. Consequently, we redirect our attention towards evaluating the potential of non-parametric spectral analysis in tackling significant structural inversion problems, alongside the mathematical obstacles involved and suggested potential solutions to overcome them.  

\subsection{Extraction of ODF from Scattering Signatures of Objects without Axial Symmetry}

Thus far, the discussion has revolved around elongated objects characterized by the inherent axial symmetry of their geometric shapes. Consequently, it is always feasible, mathematically, to ascertain a principal axis on a given projection plane of $I(\textbf{Q})$ exhibiting axial symmetry. This facilitates the inversion of the ODF conveniently by employing spherical harmonic functions with zero order. 

\begin{figure}
\centerline{
  \includegraphics[width=\linewidth]{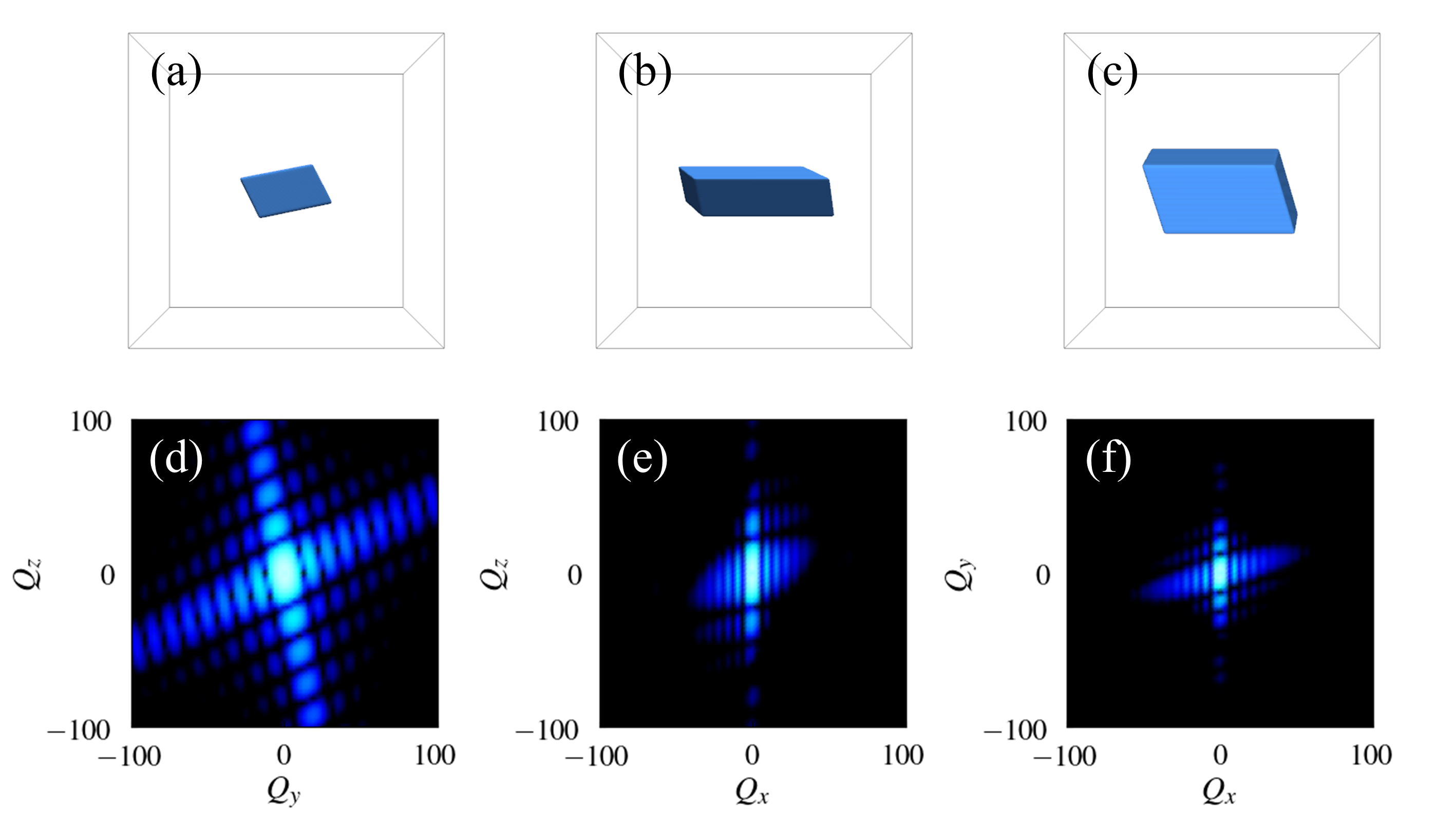}
}  
\caption{The relationship between the real space orientation (a-c) of a triclinic parallelepiped and the anisotropy of its scattering in reciprocal space (d-f) is depicted. The incident radiation source is directed towards the paper's surface. It's evident from the spectra in (d-f) that they exhibit only point symmetry.
}  
\label{fig:14}
\end{figure}

For flowing objects with shapes that lack inherent geometric symmetry, the only symmetry present in the collected scattering is point symmetry \cite{Friedel1913,Dresselhaus}, expressed as $I_p(\textbf{Q}) = I_p(-\textbf{Q})$. This spectral symmetry arises as a consequence of the two-point correlation nature of scattering intensity assuming Friedel's law \cite{Friedel1913}. Fig.~\ref{fig:14} presents an example of a triclinic parallelepiped. In this parallelepiped structure, three unequal axes intersect at non-perpendicular angles. Consequently, it lacks any mirror symmetric planes. In Panels (d-f), the projected scattering intensities display only point symmetry without any principal axis, thereby impeding the simplification of the inversion analysis of the ODF. To proper describe the orientation of non-axially symmetrical particles, the ODF should depend not only on $\Omega$ but also on $\varphi$, as depicted in Fig.~\ref{fig:1}. Due to the asymmetry, rotation about $x'$ is no longer equivalent, necessitating an additional degree of freedom to describe particle orientation. 

The average scattering intensity over an ODF $f(\Omega,\varphi)$ can be calculated using the following integral:
\begin{eqnarray}
I(\textbf{Q}) = \iint \mathrm{d}\Omega\mathrm{d}\varphi P(\textbf{Q},\Omega,\varphi) f(\Omega,\varphi),
\label{eq:4a1}
\end{eqnarray}
where $P(\textbf{Q},\Omega,\varphi)$ represents the intra-particle structure factor at a specific orientation of $\Omega$ and $\varphi$. As a result, real spherical harmonics of two angular variables cannot fully expand $f(\Omega,\varphi)$ as a function of three angular variables. In this scenario, Wigner D functions \cite{Arfken} of three angular variables as eigenfunctions should be used to expand the ODF \cite{Varshalovich1988,vanGurp}. For an axially symmetrical ODF, the arguments of the ODF can be reduced to two, allowing the ODF to be expanded using real spherical harmonics. However, the lack of axial symmetry in particle shape complicates spectral inversion analysis as spherical harmonic functions with $m \neq 0$ are necessary to reconstruct the ODF. Additionally, this lack of axial symmetry presents significant challenges in sample environment development, as multiple $I_p(\textbf{Q})$ measurements are expected to be necessary to establish sufficient boundaries for uniquely determining the ODF.

\subsection{Extraction of ODF from Scattering Signatures of Flexible Flowing Objects}

Advancements in sample environments \cite{Murphy2020} have led to the availability of flow cells capable of imparting mechanical energy comparable to the bending energy of semiflexible polymer chains immersed in solutions. Structural study of these objects is intriguing from a mathematical standpoint, as it involves the dual complexity arising from the deformation of polymers due to their flexible conformation and the orientation of stretched polymers. The potential applications of flowing polymers further contribute to the importance of this subject.

Here we choose a self-avoiding chain subjected to simple shear flow in Monte Carlo simulations as an example. Following the formalism proposed by Hsu and Binder \cite{Binder}, the Hamiltonian governing this system, denoted as $\mathbb{H}$, can be expressed as: 
\begin{eqnarray}
\mathbb{H} &=&
\frac{\kappa}{2}\int_0^L \left(\frac{\partial ^2 \textbf{r}(s)}{\partial s^2} \right)^2 ds - f\int_0^L y \frac{\partial x(s)}{\partial s} ds \nonumber\\
&=& \sum_{i}E_{\mathrm{bend}}(\theta_i) - E_{\mathrm{shear}}(\textbf{r}_i).
\label{eq:4.a}
\end{eqnarray}
The first term of Eqn.~\eqref{eq:4.a} represents the intrinsic bending energy for a self-avoiding chain with a contour length of $L$ consisting of all $i$ segments. $\kappa$ denotes a constant associated with bending energy. $\textbf{r}(s)$ represents the segmental direction tangent to the contour path $s$. The second term represents the competing shear energy, where $f$ is the magnitude of the applied mechanical force. $x$ and $y$ denote the positional variables for segments. Eqn.~\eqref{eq:4.a} holds in the continuous limit. For the discrete case, it can be reformulated as:
\begin{equation}
\mathbb{H} = \sum_{i}\epsilon \cdot \frac{\theta_i^2}{2} - \gamma \cdot r_i^y \frac{\partial r_i^x(s)}{\partial s}.
\label{eq:4.b}
\end{equation}
Here $\epsilon$ can be considered as the bending modulus. $\theta_i$ is the the angle between the $i^{th}$ segment with its neighboring segments. $\gamma$ is the shear strain.

\begin{figure}
\centerline{
  \includegraphics[width=\linewidth]{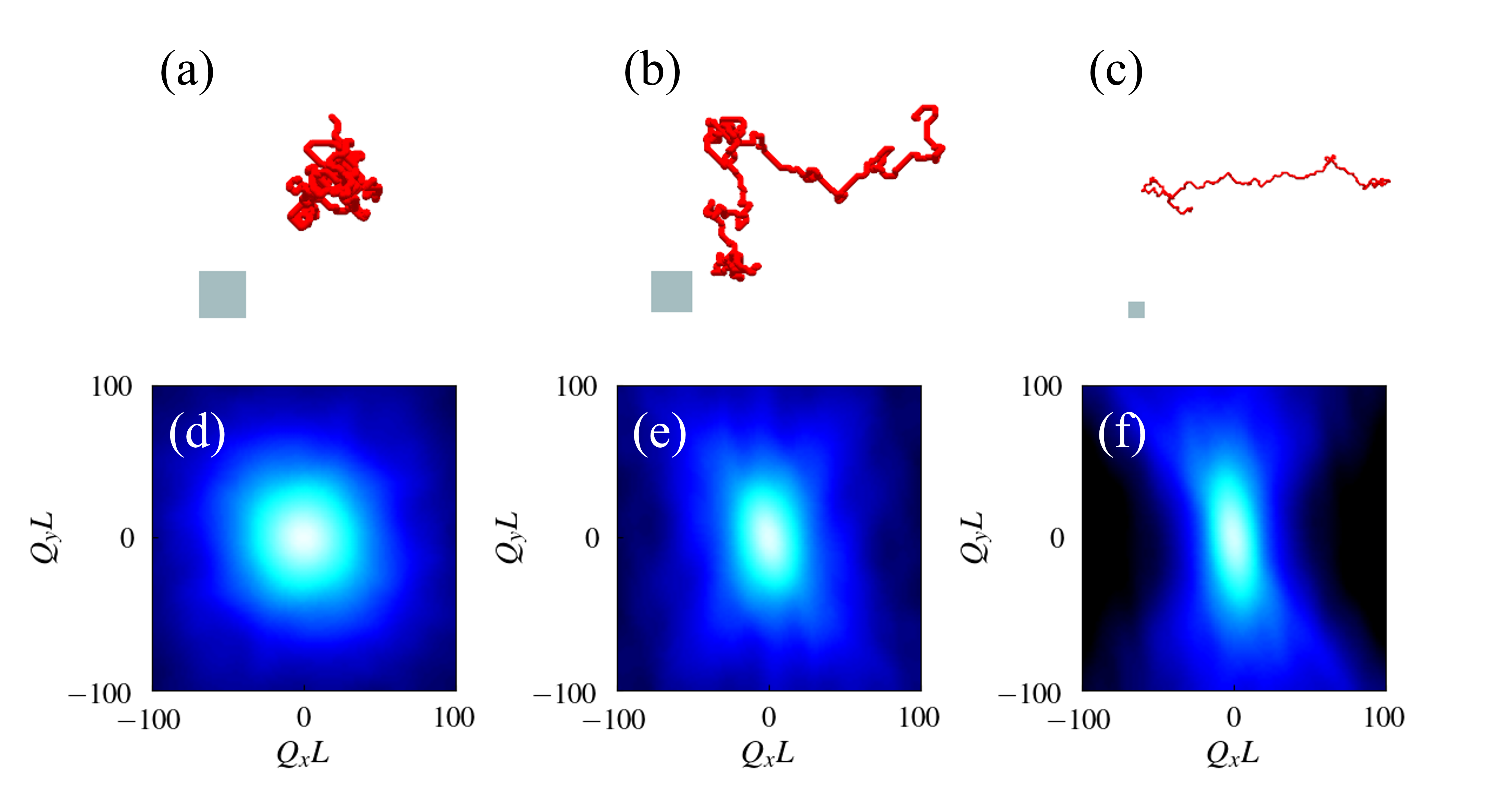}
}  
\caption{The real-space configurations of a flexible self-avoiding chain at its quiescent state (a), and under two different shear energies of $0.01$ $k_BT$ and $0.04$ $k_BT$, are illustrated in (b) and (c) respectively. The gray squares in each panel denote the same length scale. Corresponding scattering spectra acquired from the velocity-velocity gradient ($xy$) planes are presented in (d)-(f). Here, $L$ represents the contour length of the self-avoiding chain.
}  
\label{fig:15}
\end{figure}

When the applied shear energy approaches the magnitude of the intrinsic bending energy, the system can be substantially driven away from its equilibrium configurations as displayed in Fig.~\ref{fig:15}(a-c). This structural distortion becomes apparent through the analysis of $I_{xy}(\textbf{Q})$, representing the anisotropy of projection of $I(\textbf{Q})$ onto the flow-gradient plane, as illustrated in Fig.~\ref{fig:15}(d-f). 

\begin{figure}
\centerline{
  \includegraphics[scale = 0.5]{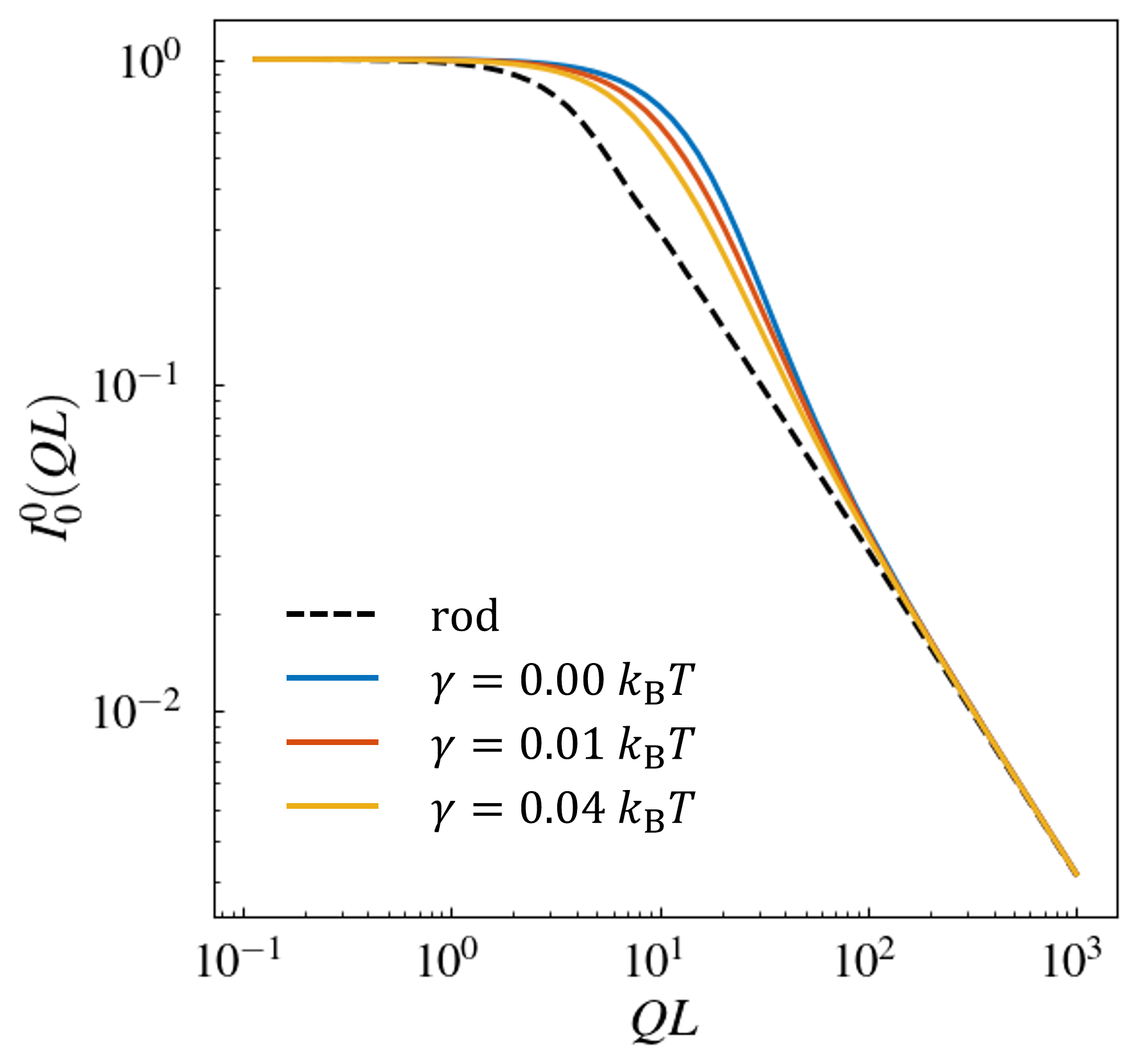}
}  
\caption{The isotropic component of $I(\textbf{Q})$ for a self-avoiding chain under steady shear, characterized by various strain rates, is derived from the analysis of $I_{xy}(\textbf{Q})$ depicted in Fig.~\ref{fig:15}(d-f). The dashed line in the figure corresponds to the asymptotic scattering function of a rigid rod with a length $L$ equivalent to the contour length of the self-avoiding chain. The results are presented in dimensionless units, scaled by $QL$.
}  
\label{fig:16}
\end{figure}

The contrasting deformation behaviors between a shear self-avoiding chain and the rotation of a rigid body are distinctly discernible through the isotropic component of $I(\textbf{Q})$, denoted as $I_0^0(Q)$, as obtained from the spectral analysis of $I_{xy}(\textbf{Q})$, with the findings presented in Fig.~\ref{fig:16}.  

In both the low $Q$ region ($QL < 1$) and the high $Q$ region ($QL > 100$), their behavior closely resembles that of the reference rigid rod system with a contour length $L$, suggesting that the contour length of the self-avoiding chain remains consistent within the investigated range of applied shear stress. However, a characteristic evolution of $I_0^0(Q)$ is observed in the range $1 < QL < 100$. This observation clearly reveals that for a flowing self-avoiding chain, the deformation not only involves alignment similar to flowing rigid elongated objects, but also entails alterations in the angular correlation of the constituent segments due to the applied shear energy. 

Referring to the extracted $I_0^0(QL)$ depicted in Fig.~\ref{fig:16}, one deduces a significant feature of flowing polymers: the dependence of their ODF on $Q$. Indeed, similar results have been in deformed polymer melts \cite{wang2017fingerprinting,Lam2018Scaling,wang2020quantitative,sun2021molecular,lam2023quantifying}. This feature clearly suggests the necessity of addressing an additional angular degree of freedom to accurately reconstruct the configurations of flowing flexible objects from their scattering signatures. While the information of ODF at larger length scales can be obtained through generalized Guinier analysis in the low $Q$ region \cite{HuangJPCL, lam2023quantifying}, it is expected that extracting the ODF at smaller length scales will involve more intricate mathematical procedures.  

\subsection{Extraction of ODF from Scattering Signatures of Flowing Objects with Size Polydispersity}

Up to this point, our discussion of data analysis has operated under the assumption that the dispersed particles are of uniform size. However, in reality, both naturally occurring and synthetic soft materials often present a continuous distribution of constituent sizes.

Experimentally measured scattering intensities obscure information about particle size distribution. The particle size distribution can generally be expressed through a real-space three-dimensional function $f_s(\vec{R})$, where $\vec{R}$ represents the particle size along different dimensions. By incorporating $f_s(\vec{R})$ to address size polydispersity, Eqn.~\eqref{eq:2.1} can be extended as:
\begin{eqnarray}
    I({\bf Q}) = \iint \mathrm{d}\Omega \mathrm{d} \Vec{R} P({\bf Q}, \Omega,\Vec{R}) F(\Omega, \Vec{R}),
    \label{eq:4.c}
\end{eqnarray}
In this context, $P(\mathbf{Q}, \Omega,\vec{R})$ represents the particle form factor at a specific orientation and size, while $F(\Omega, \vec{R})$ denotes the probability distribution of locating particles at orientation $\Omega$ and size $\vec{R}$. Evidently, the introduction of polydispersity complicates the inversion of the ODF, as it couples the particle orientation with the size distribution via $F(\Omega, \vec{R})$, which typically lacks a known mathematical expression. 

Similar to the methodology used in addressing spectral analysis of interacting polydispersed systems \cite{Beta1, Beta2, Chen1986}, a convenient tactic involves factorizing $F(\Omega, \vec{R})$ as $f(\Omega)f_s(\vec{R})$. However, the limitation of this decoupling approximation for spectral analysis of polydispersed systems at quiescent states has been long recognized \cite{Pusey}. Assessing the applicability of this approach to extract $f(\Omega)$ for non-equilibrium states, while accounting for both size polydispersity and size geometry, undoubtedly demands thorough computational benchmarking. 

With moderate size polydispersity, it may be mathematically feasible to accurately determine both $f(\Omega)$ and $f_s(\vec{R})$ by employing an appropriate combination of real spherical harmonic basis expansion for $f(\Omega)$ and central moment expansion for $f_s(\vec{R})$ \cite{Huang2023b}. However, in systems with substantial size distribution, the assumptions underlying the expansion scheme may not hold, as it involves expressing moments of a random variable around its mean in terms of moments of the deviations from the mean. In such scenarios, as the connection between the ODF and collected scattering spectra involves many complicated degrees of freedom, alternative methods such as machine learning, supported by extensive computational simulations, may offer a more suitable approach for inverting the ODF.   

\subsection{Addressing the Effect of Instrument Resolution}

Due to variations in the wavelength distribution of incoming neutrons, optical collimation, and detector sensitivity, the collected intensity of small angle neutron scattering inevitably experiences smearing. To our knowledge, there has not been any attempt to apply two-dimensional desmearing techniques to alleviate the effects of instrument resolution on the scattering of moving objects, with the objective of correcting potential errors in the determination of $f(\Omega)$. 

The relevance of this implementation to related scattering studies can be illustrated by the following example of rod-like micelles, which have been a focal point in research aimed at investigating the ODF of elongated particles under steady shear flows initiated by Hayter and coworkers \cite{Hayter1984}. However, when similar rod-like micelles are studied at different scattering facilities, distinct mathematical expressions of the ODF are derived through trial-and-error parametric analytical modeling \cite{Hayter1984, Foerster2005}. By properly analyzing desmeared two-dimensional spectra, it becomes possible to unambiguously determine whether the observed discrepancy arises from subtle chemical differences or from biased information blurred by variations in instrument resolutions across different facilities. 

\begin{figure}
\centerline{
  \includegraphics[width=\linewidth]{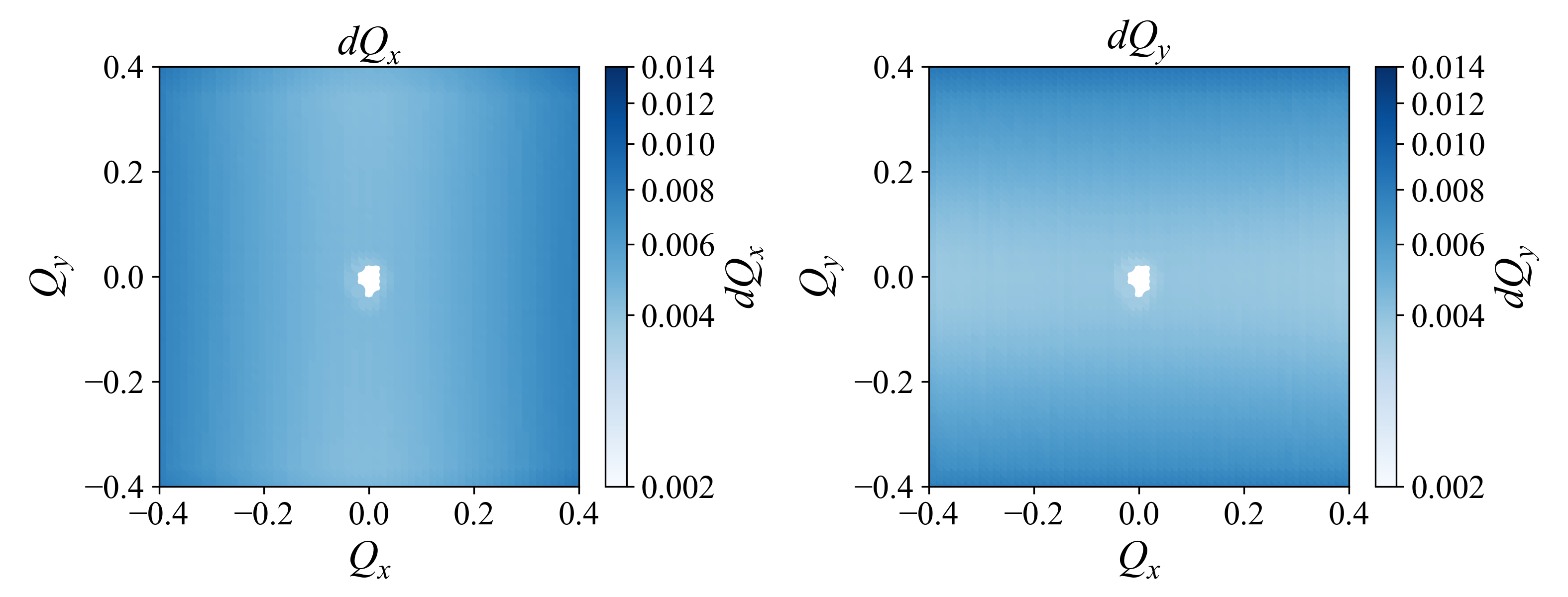}
}  
\caption{Resolution maps depicting (a) $Q_x$ and (b) $Q_y$ for the Extended Q-Range Small Angle Neutron Scattering Diffractometer (EQSANS) at the Spallation Neutron Source, Oak Ridge National Laboratory. The standard deviations of the instrumental resolution function along the $Q_x$ and $Q_y$ directions are denoted by $\mathrm{d}Q_x$ and $\mathrm{d}Q_y$, respectively. The calculations are based on incident neutrons with a wavelength band of $2.6~\text{\AA} - 3.6~\text{\AA}$ at 2.5~m sample to detector distance.   
}  
\label{fig:17}
\end{figure}

Similar to the previous section, the combined effect of instrumental resolution and particle orientation can be modeled by the following multiple integral:
\begin{eqnarray}
    I(\textbf{Q}) = \iint \mathrm{d}\Omega\mathrm{d}\textbf{r} P(\textbf{r},\textbf{Q}_{\perp},\Omega) R(\textbf{r}-\textbf{Q}_{\parallel}) f(\Omega),
    \label{eq:4.d}
\end{eqnarray}
where $R(\textbf{r})$ is the resolution function, $\textbf{r}$ is a two-dimensional vector on the detector plane and $\textbf{Q}_{\parallel}$ and $\textbf{Q}_{\perp}$ are the component of $\textbf{Q}$ parallel and perpendicular to $\textbf{r}$, respectively. Eqn.~\eqref{eq:4.d} illustrates that how instrumental resolution affects the quantitative extraction of the ODF from scattering data. With an isotropic resolution function, the convoluted integral only modifies the isotropic component of $I_p(\textbf{Q})$. However, as shown in Fig.~\ref{fig:17}, the instrumental resolution varies along horizontal and vertical directions, leading to an symmetrical variance distribution of $R(\textbf{Q}_{\parallel})$, which is characterized by the standard deviations of $\mathrm{d}Q_x$ and $\mathrm{d}Q_y$ along $Q_x$ and $Q_y$ directions over the $Q_{x}$ and $Q_{y}$ plane. As a result, the anisotropic components of the measured $ I_p(\textbf{Q})$ collectively manifests the contribution from the anisotropic components of $R(\textbf{Q}_{\parallel})$ and the ODF of flowing particles. This inevitably affects the quantitative determination of the scattering order parameter $\hat{S}_l^m(Q)$ and, consequently, the accurate reconstruction of the ODF.

To ensure optimal determination of $f(\Omega)$ from $I_p(\textbf{Q})$, it is critical to address the impact of instrument resolution beforehand, thus enhancing the accuracy of spectral data analysis. One viable approach involves extending the central moment expansion methodology of the desmearing technique \cite{Huang2023} to two dimensions. As the variance of $R(\textbf{Q}_{\parallel})$ for currently operational SANS instruments typically remains sufficiently constrained, the convergence of the expansion scheme can be ensured. 

\section{Concluding Remarks}

We conclude this commentary with two pivotal thoughts. Firstly, after reviewing the advantages and limitations of existing parametric and non-parametric methods for extracting the ODF of flowing objects, the versatility of the non-parametric approach becomes apparent: by selecting an appropriate coordinate system and basis functions that align with the configuration and orientation of flowing elongated objects in real space, while also preserving the compatibility with the deformation gradient of the applied mechanical condition, the ODF can be extracted without bias from the spectral anisotropy. This advantage stands unmatched by the parametric approaches, which encounter insurmountable difficulties in deriving the exact expression of the ODF for different flow conditions.

Secondly, our literature review suggests that both existing parametric and non-parametric approaches may fall short in providing sufficient quantitative insights into the alignment behavior of general soft materials under diverse deformation conditions. It is evident that a new spectrum of analytical strategies is necessary to effectively address this gap. Further advancements can be anticipated through the utilization of machine learning methods. Several approaches have been developed to tackle the challenging spectral inversion problems of soft materials in their quiescent states \cite{Chang2022, Tung2022, Tung2024}, which cannot be achieved through analytical analysis alone. The potential extensions of existing machine learning-based spectral analysis approaches hold promise in addressing the intricate structural inversion challenges encountered in highly interactive soft matter systems, such as concentrated polymer or polyelectrolyte solutions, under various flow conditions.     

\clearpage







\ack{Acknowledgements}

A portion of this research used resources at the Spallation Neutron Source and the Center for Nanophase Materials Sciences, DOE Office of Science User Facilities operated by the Oak Ridge National Laboratory. This research was sponsored by the Laboratory Directed Research and Development Program of Oak Ridge National Laboratory, managed by UT-Battelle, LLC, for the U. S. Department of Energy. G.R.H. is supported by the National Science and Technology Council (NSTC) in Taiwan with Grant No. NSTC 111-2112-M-110-021-MY3. YS was supported by the U.S. Department of Energy, Office of Science, Office of Basic Energy Sciences, Materials Science and Engineering Division. RPM acknowledges support from CHRNS, a national user facility jointly funded by the NCNR and the NSF under Agreement No. DMR-2010792. Commercial equipment or software identified in this work does not imply recommendation nor endorsement by NIST. YW acknowledges the support by the U.S. Department of Energy, Office of Science, Office of Basic Energy Sciences, Early Career Research Program Award KC0402010, under Contract DE-AC05-00OR22725.

\clearpage

\bibliographystyle{iucr}
\bibliography{references}

@book{Schiff,
  author = {Schiff, L. I.},
  title  =  {Quantum Mechanics},
  publisher = {McGraw-Hill},
  address={New York},
  year   = {1968}       
}

@article{Gantenbein2018,
  title={Three-dimensional printing of hierarchical liquid-crystal-polymer structures},
  author={Gantenbein, S. and Masania, K. and Woigk, W. and Sesseg, J. P. W. and Tervoort, T. A. and Studart, A. R.},
  journal={Nature},
  volume={561},
  pages={226--230},
  year={2018},
  publisher={Nature Publishing Group UK London}
}

@article{Fratzl2007,
  title={Nature’s hierarchical materials},
  author={Fratzl, Peter and Weinkamer, Richard},
  journal={Prog. Mater. Sci.},
  volume={52},
  pages={1263--1334},
  year={2007},
  publisher={Elsevier}
}

@article{Long2012,
  title={Recent advances in large-scale assembly of semiconducting inorganic nanowires and nanofibers for electronics, sensors and photovoltaics},
  author={Long, Y.-Z. and Yu, M. and Sun, B. and Gu, C.-Z. and Fan, Z.},
  journal={Chem. Soc. Rev.},
  volume={41},
  pages={4560--4580},
  year={2012},
  publisher={Royal Society of Chemistry}
}

@article{Blell2017,
  title={Generating in-plane orientational order in multilayer films prepared by spray-assisted layer-by-layer assembly},
  author={Blell, R. and Lin, X. and Lindstrom, T. and Ankerfors, M. and Pauly, M. and Felix, O. and Decher, G.},
  journal={ACS Nano},
  volume={11},
  pages={84--94},
  year={2017},
  publisher={ACS Publications}
}

@article{Richard2013,
  title={Molecular orientation in electrospun fibers: from mats to single fibers},
  author={Richard-Lacroix, M. and Pellerin, C.},
  journal={Macromolecules},
  volume={46},
  pages={9473--9493},
  year={2013},
  publisher={ACS Publications}
}

@article{Patil2010,
  title={A study on the chain-particle interaction and aspect ratio of nanoparticles on structure development of a linear polymer},
  author={Patil, N. and Balzano, L. and Portale, G. and Rastogi, S.},
  journal={Macromolecules},
  volume={43},
  pages={6749--6759},
  year={2010},
  publisher={ACS Publications}
}

@article{Wu2017,
  title={Shear flow induced long-range ordering of rod-like viral nanoparticles within hydrogel},
  author={Wu, Y. and Jiang, Z. and Zan, X. and Lin, Y. and Wang, Q.},
  journal={Colloids Surf. B: Biointerfaces},
  volume={158},
  pages={620--626},
  year={2017},
  publisher={Elsevier}
}

@article{Roe,
  title={Description of Crystallite Orientation in Polycrystalline Materials Having Fiber Texture},
  author={Roe, R.-J. and Krigbaum, W. R.},
  journal={J. Chem. Phys.},
  volume={40},
  pages={2608--2615},
  year={1964}
}

@article{Lovell,
  title={Molecular orientation distribution derived from an arbitrary reflection},
  author={Lovell, R. and Mitchell, G. R.},
  journal={Acta Crvst. A},
  volume={37},
  pages={135--137},
  year={1981}
}

@article{vanGurp,
  author={van Gurp, M.},
  journal={Colloid Polym. Sci.},
  volume={273},
  pages={607--625},
  year={1995}
}

@article{Onsager,
  author={Onsager, L.},
  journal={Ann. N. Y. Acad. Sci.},
  volume={51},
  pages={627--659},
  year={1949}
}

@article{Franco2008,
  title={A generalisation of the Onsager trial-function approach: describing nematic liquid crystals with an algebraic equation of state},
  author={Franco-Melgar, M. and Haslam, A. J. and Jackson, G.},
  journal={Mol. Phys.},
  volume={106},
  pages={649--678},
  year={2008},
  publisher={Taylor \& Francis}
}

@article{Franco2009,
  title={Advances in generalised van der Waals approaches for the isotropic--nematic fluid phase equilibria of thermotropic liquid crystals--an algebraic equation of state for attractive anisotropic particles with the Onsager trial function},
  author={Franco-Melgar, M. and Haslam, A. J. and Jackson, G.},
  journal={Mol. Phys.},
  volume={107},
  pages={2329--2358},
  year={2009},
  publisher={Taylor \& Francis}
}

@article{Binder,
  title={Stretching semiflexible polymer chains: Evidence for the importance of excluded volume effects from Monte Carlo simulation},
  author={Hsu, H.-P. and Binder, K.},
  journal={J. Chem. Phys.},
  volume={136},
  pages={024901},
  year={2012}
}

@article{Helgeson2024,
  title={Self-Consistent Connected-Rod Model for Small-Angle Scattering from Deformed Semiflexible Chains in Flow},
  author={Zhang, Z. and Smith, G. S. and Corona, P. T. and Underhill, P. T. and Leal, L. G. and Helgeson, M. E.},
  journal={Macromolecules},
  volume={57},
  pages={201--216},
  year={2024}
}

@article{Chen1986,
  title={Small Angle Neutron Scattering Studies of the Structure and Interaction in Micellar and Microemulsion Systems},
  author={Chen, S.-H.},
  journal={Ann. Rev. Phys. Chem.},
  volume={37},
  pages={351--399},
  year={1986}
}

@incollection{Pusey,
  author    = {Pusey, P.},
  title     = {Colloidal Suspensions},
  publisher = {North Holland},
  address={Amsterdam},
  year={1991},
  booktitle = {Liquids, Freezing and the Glass Transition, Ed. by J.-P. Hansen, D. Levesque, J. Zinn-Justin}
}

@article{Wagner1996,
  title={SANS Analysis of the Molecular Order in Poly(\gamma-benzylL-glutamate)/Deuterated Dimethylformamide (PBLG/d-DMF) under Shear and during Relaxation},
  author={Walker, L. M. and Wagner, N. J.},
  journal={Macromolecules},
  volume={29},
  pages={2298--2301},
  year={1996}
}

@article{Tabor2022,
  title={Exploring shear alignment of concentrated wormlike micelles using rheology coupled with small-angle neutron scattering},
  author={King, J. P. and Butler, C. S. G. and Prescott, S. W. and Sokolova, A. V. and de Campo, L. and Williams, A. P. and Tabor, R. F.},
  journal={Phys. Fluids},
  volume={34},
  pages={083104},
  year={2022}
}

@article{Hayter1984,
  title={Use of Viscous Shear Alignment To Study Anisotropic Micellar Structure by Small-Angle Neutron Scattering},
  author={Hayter, J. B. and Penfold, J.},
  journal={J. Phys. Chem.},
  volume={88},
  pages={4589-4593},
  year={1984}
}

@article{Scheraga1951,
  title={Double refraction of flow: Numerical evaluation of extinction angle and birefringence as a function of velocity gradient},
  author={Scheraga, H. A. and Edsall, J. T. and Gadd Jr., J. O.},
  journal={J. Chem. Phys.},
  volume={19},
  pages={1101--1108},
  year={1951},
  publisher={American Institute of Physics}
}

@article{Jerrard1959,
  title={Theories of streaming double refraction},
  author={Jerrard, H. G.},
  journal={Chem. Rev.},
  volume={59},
  pages={345--428},
  year={1959},
  publisher={ACS Publications}
}

@article{Foerster2005,
  title={Shear Thinning and Orientational Ordering ofWormlike Micelles},
  author={Förster, S. and Konrad, M. and Lindner, P.},
  journal={Phys. Rev. Lett.},
  volume={88},
  pages={4589-4593},
  year={2005}
}

@article{Haywood2017,
  title={New Insights into the Flow and Microstructural Relaxation Behavior of Biphasic Cellulose Nanocrystal Dispersions from RheoSANS},
  author={Haywood, A. D. and Weigandt, K. M. and Saha, P. and Noor, M. and Green, M. J. and Davis, V. A.},
  journal={Soft Matter},
  volume={13},
  pages={8451-8462},
  year={2017}
}

@article{Huang2019,
  title={Orientational Distribution Function of Aligned Elongated Molecules and Particulates Determined from Their Scattering Signature},
  author={Huang, G.-R. and Wang, Y. and Do, C. and Shinohara, Y. and Egami, T. and Porcar, L. and Liu, Y. and Chen, W.-R.},
  journal={ACS Macro Lett.},
  volume={8},
  pages={1257-1262},
  year={2019}
}

@article{Huang2021,
  title={An Exact Inversion Method for Extracting Orientation Ordering by Small-Angle Scattering},
  author={Huang, G.-R. and Carrillo, J. M. and Wang, Y. and Do, C. and Porcar, L. and Sumpter, B. and Chen, W.-R},
  journal={Phys. Chem. Chem. Phys.},
  volume={23},
  pages={4120–4132},
  year={2021}
}

@article{Lang2016,
  title={The connection between biaxial orientation and shear thinning for quasi-ideal rods},
  author={Lang, C. and Kohlbrecher, J. and Porcar, L. and Lettinga, M. P.},
  journal={Polymers},
  volume={8},
  pages={291},
  year={2016},
  publisher={MDPI}
}

@article{Lang2019,
  title={Microstructural Understanding of the Length- and Stiffness-Dependent Shear Thinning in Semidilute Colloidal Rods},
  author={Lang, C. and Kohlbreche, J. and Porcar, L. and  Radulescu A. and S. Karin and Dhont, J. K. G. and Lettinga, M. P. },
  journal={Macromolecules},
  volume={52},
  pages={9604-9612},
  year={2019}
}

@article{Li2020,
  title={Continuous crystalline graphene papers with gigapascal strength by intercalation modulated plasticization},
  author={Li, P. and Yang, M. and Liu, Y. and Qin, H. and Liu, J. and Xu, Z. and Liu, Y. and Meng, F. and Lin, J. and Wang, F. and Gao, C.},
  journal={Nat. Commun.},
  volume={11},
  pages={2645},
  year={2020}
}

@article{Munier2022,
  title={Rheo-SAXS study of shear-induced orientation and relaxation of cellulose nanocrystal and montmorillonite nanoplatelet dispersions},
  author={Munier, P. and Hadi, S. E. and Segad, M. and Bergstr\"om, L.},
  journal={Soft Matter},
  volume={18},
  pages={390–396},
  year={2022}
}

@article{Huang2023,
  title={Desmearing small-angle scattering data by central moment expansions},
  author={Huang, G.-H. and Tung, C.-H. and Chen, M.-Z. and Porcar, L. and Shinohara, Y. and Wildgruber, C. U. and Do, C. and Chen, W.-R.},
  journal={J. Appl. Cryst.},
  volume={56},
  pages={1537–1543},
  year={2023}
}

@article{Huang2023b,
  title={ Model-Free Approach for Profiling of Polydisperse Soft Matter Using Small Angle Scattering},
  author={Huang, G.-R. and Tung, C.-H. and Porcar, L. and Wang, Y. and Shinohara, Y. and Do, C. and Chen, W.-R.},
  journal={Macromolecules},
  volume={56},
  pages={6436--6443},
  year={2023}
}

@book{Varshalovich1988,
    author = {Varshalovich, D. A. and Moskalev, A. N. and Khersonskii, V. K.},
    title = {Quantum Theory of Angular Momentum},
    publisher = {World Scientific},
    address={Singapore},
    year = {1988}
}

@article{Okagawa1973,
  title={The kinetics of flowing dispersions. VI. Transient orientation and rheological phenomena of rods and discs in shear flow},
  author={Okagawa, A. and Cox, R. G. and Mason, S. G.},
  journal={J. Colloid Interface Sci.},
  volume={45},
  pages={303--329},
  year={1973},
  publisher={Elsevier}
}

@article{Maier1959,
  title={Eine einfache molekular-statistische Theorie der nematischen kristallinfl{\"u}ssigen Phase. Teil l1.},
  author={Maier, W. and Saupe, A.},
  journal={Z. Naturforsch A},
  volume={14},
  pages={882--889},
  year={1959},
  publisher={Verlag der Zeitschrift f{\"u}r Naturforschung}
}

@article{Maier1960,
  title={Eine einfache molekular-statistische Theorie der nematischen kristallinfl{\"u}ssigen Phase. Teil II},
  author={Maier, W. and Saupe, A.},
  journal={Z. Naturforsch A},
  volume={15},
  pages={287--292},
  year={1960},
  publisher={Verlag der Zeitschrift f{\"u}r Naturforschung}
}

@book{De1993,
  title={The Physics of Liquid Crystals},
  author={De Gennes, P.-G. and Prost, J.},
  year={1993},
  publisher={Oxford University Press},
  address={Oxford},
}

@article{Lemaire2002,
  title={The measurement by SAXS of the nematic order parameter of laponite gels},
  author={Lemaire, B. J. and Panine, P. and Gabriel, J.-C. P. and Davidson, P.},
  journal={Europhysics Letters},
  volume={59},
  pages={55},
  year={2002},
  publisher={IOP Publishing}
}

@book{Rubinstein2003,
  title={Polymer Physics},
  author={Rubinstein, M. and Colby, H. C.},
  year={2003},
  publisher={Oxford University Press},
  address={Oxford}
}

@article{Hakansson2016,
  title={Nanofibril Alignment in Flow Focusing: Measurements and Calculations},
  author={Håkansson, K. M. O. and Lundell, F. and Prahl-Wittberg, L. and S\"oderberg, L. D.},
  journal={J. Phys. Chem. B},
  volume={120},
  year={2016},
  pages={6674--6686},
  doi={10.1021/acs.jpcb.6b01983}
}

@article{Jeffery1922,
  title={The motion of ellipsoidal particles immersed in a viscous fluid},
  author={Jeffery, G. B.},
  journal={Proc. R. Soc. London Ser. A},
  volume={102},
  pages={161--179},
  year={1922},
  publisher={The Royal Society London}
}

@article{Batchelor1970,
  title={The stress system in a suspension of force-free particles},
  author={Batchelor, G. K.},
  journal={J. Fluid Mech.},
  volume={41},
  pages={545--570},
  year={1970},
  publisher={Cambridge University Press}
}

@article{Hinch1976,
  title={Constitutive equations in suspension mechanics. Part 2. Approximate forms for a suspension of rigid particles affected by Brownian rotations},
  author={Hinch, E. J. and Leal, L. G.},
  journal={J. Fluid Mech.},
  volume={76},
  pages={187--208},
  year={1976},
  publisher={Cambridge University Press}
}

@article{Luckhurst1977,
  title={Why Is the Maier–Saupe Theory of Nematic Liquid Crystals so Successful?},
  author={Luckhurst, G. R. and Zannoni, C.},
  journal={Nature},
  volume={267},
  pages={412–414},
  year={1977}
}

@book{Doi1986,
  title={The Theory of Polymer Dynamics},
  author={Doi, M. and Edwards, S. F. },
  year={1986},
  publisher={Clarendon Press},
  address={Oxford}
}

@book{Larson2013,
  title={Constitutive equations for polymer melts and solutions: Butterworths series in chemical engineering},
  author={Larson, R. G.},
  year={2013},
  publisher={Butterworth-Heinemann}
}

@book{Bird1987,
  title={Dynamics of polymeric liquids, volume 2: Kinetic theory},
  author={Bird, R. B. and Curtiss, C. F. and Armstrong, R. C. and Hassager, O.},
  year={1987},
  publisher={Wiley}
}

@article{Wu2014,
  title={Understanding and Describing the Liquid-Crystalline States of Polypeptide Solutions: A Coarse-Grained Model of PBLG in DMF},
  author={Wu, L. and M\"uller, E. A. and Jackson, G. },
  journal={Macromolecules},
  volume={47},
  pages={1482–1493},
  year={2014}
}

@article{Hayter1987,
  title={A Small-angle Neutron Scattering Investigation of Rod-like Micelles aligned by Shear Flow},
  author={Cummins, P. G. and Staples, E. and Hayter, J. B. and Penfold, J.},
  journal={J. Chem. Soc., Faraday Trans. 1},
  volume={83},
  pages={2773–2786},
  year={1987}
}

@article{Penfold1988,
  title={Small-Angle Neutron Scattering Studies of Systems Undergoing Shear},
  author={Penfold, J.},
  journal={J. Appl. Cryst.},
  volume={21},
  pages={770–776},
  year={1988}
}

@article{Penfold1991,
  title={SMALL ANGLE NEUTRON SCATTERING INVESTIGATION OF RODLIKE MICELLES ALIGNED BY SHEAR FLOW},
  author={Penfold, J. and Staples, E. and Cummins, P. G.},
  journal={Adv. Colloid Interface Sci.},
  volume={34},
  pages={451–476},
  year={1991}
}

@article{Herbst1986,
  title={Orientational relaxation of aligned rod-like micelles on a time scale of 300 ms},
  author={Herbst, L. and Hoffmann, H. and Kalus, J. and Thurn, H. and Ibel, K. and May, R. P.},
  journal={Chem. Phys.},
  volume={103},
  pages={437--445},
  year={1986},
  publisher={Elsevier}
}

@article{Kalus1988,
  title={Transient SANS studies of rodlike micelles on a time scale of 100 ms},
  author={Kalus, J. and Chen, S. H. and Hoffmann, H. and Neubauer, G. and Lindner, P. and Thurn, H.},
  journal={J. Appl. Cryst.},
  volume={21},
  pages={777--780},
  year={1988},
  publisher={International Union of Crystallography}
}

@article{Bihannic2010,
  title={Orientational Order of Colloidal Disk-Shaped Particles under Shear-Flow Conditions: a Rheological-Small-Angle X-ray Scattering Study},
  author={Bihannic, I. and Baravian, C. and Duval, J. F. L. and Paineau, E. and Meneau, F. and Levitz, P. and de Silva, J. P. and Davidson, P. and Michot, L. J.},
  journal={J. Phys. Chem. B},
  volume={114},
  pages={16347–16355},
  year={2010}
}

@article{Bihannic2011,
  title={Rheo-SAXS investigation of shear-thinning behaviour of very anisometric repulsive disc-like clay suspensions},
  author={Philippe, A. M. and Baravian, C. and Imperor-Clerc, M. and de Silva, J. and Paineau E. and Bihannic, I. and Davidson, P. and Meneau, F. and Levitz, P. and Michot, L. J.},
  journal={J. Phys.: Condens. Matter},
  volume={23},
  pages={194112},
  year={2011}
}

@article{Burghardt1996,
  title={Birefringence, X-ray Scattering, and Neutron Scattering Measurements of Molecular Orientation in Sheared Liquid Crystal Polymer Solutions},
  author={Hongladarom, K. and Ugaz, V. M. and Cinader, D. K. and Burghardt, W. R. and Quintana, J. P.},
  journal={Macromolecules},
  volume={29},
  pages={5346-5355},
  year={1996}
}

@article{Burghardt1998,
  title={In Situ X-ray Scattering Study of a Model Thermotropic Copolyester under Shear: Evidence and Consequences of Flow-Aligning Behavior},
  author={Ugaz, V. M. and Burghardt, W. R.},
  journal={Macromolecules},
  volume={31},
  pages={8474-8484},
  year={1998}
}

@article{Lindner1998,
  title={Structure and rheology of concentrated wormlike micelles at the shear-induced isotropic-to-nematic transition},
  author={Berret, J.-F. and Roux, D. C. and Lindner, P.},
  journal={Eur. Phys. J. B},
  volume={5},
  pages={67-77},
  year={1998}
}

@article{Hsiao2005,
  title={Shear-Induced Molecular Orientation and Crystallization in Isotactic Polypropylene: Effects of the Deformation Rate and Strain},
  author={Somani, R. H. and Yang, L. and Hsiao, B. S. and Sun, T. and Pogodina, N. V. and Lustiger, A.},
  journal={Macromolecules},
  volume={38},
  pages={1244-1255},
  year={2005}
}

@article{Hsiao2006,
  title={X-ray studies of regenerated cellulose fibers wet spun from cotton linter pulp in NaOH/thiourea aqueous solutions},
  author={Chen, X. and Burger, C. and Fang, D. and Ruan, D. and Zhang, L. and Hsiao, B. S. and Chu, B.},
  journal={Polymer},
  volume={47},
  pages={2839–2848},
  year={2006}
}

@article{Porcar2008,
  title={Orientational distributions and nematic order of rodlike magnetic nanoparticles in dispersions},
  author={Krishnamurthy, V. V. and Mankey, G. J. and He, B. and Piao, M. and Wiest, J. M. and Nikles, D. E. and Porcar, L. and Robertson, J. L.},
  journal={Phys. Rev. E},
  volume={77},
  pages={031403},
  year={2008}
}

@article{Porcar2010,
  title={Shear-Induced Nanometer and Micrometer Structural Responses in Nanocomposite Hydrogels},
  author={Loizou, E. and Porcar, L. and Schexnailder, P. and Schmidt, G. and Butler, P.},
  journal={Macromolecules},
  volume={43},
  pages={1041–1049},
  year={2010}
}

@article{Hakansson2014,
  title={Hydrodynamic alignment and assembly of
nanofibrils resulting in strong cellulose filaments},
  author={Håkansson, K. M. O. and Fall, A. B. and Lundell, F. and Yu, S. and Krywka, C. and Roth, S. V. and Santoro, G. and Kvick, M. and Wittberg, L. P and Wågberg, L. Söderberg, L. D.},
  journal={Nat. Commun.},
  volume={5},
  pages={4018},
  year={2014}
}

@article{Winnik2011,
  title={Probing the Structure of the Crystalline Core of Field-Aligned, Monodisperse, Cylindrical Polyisoprene-block-Polyferrocenylsilane Micelles in Solution Using Synchrotron Small- and Wide-Angle X-ray Scattering},
  author={Gilroy, J. B. and Rupar, P. A. and Whittell, G. R. and Chabanne, L. and Terrill, N. J. and Winnik, M. A. and Manners, I. and Richardson, R. M.},
  journal={J. Am. Chem. Soc.},
  volume={133},
  pages={17056–17062},
  year={2011}
}

@article{Wagner1997,
  title={In Situ Analysis of the Defect Texture in Liquid Crystal Polymer Solutions under Shear},
  author={Walker, L. M. and Kernick III, W. A. and Wagner, N. J.},
  journal={Macromolecules},
  volume={30},
  pages={508--514},
  year={1997}
}

@article{Wagner2005,
  title={The rheology and microstructure of acicular precipitated calcium carbonate colloidal suspensions through the shear thickening transition},
  author={Egres, R. G. and Wagner, N. J.},
  journal={J. Rheol.},
  volume={49},
  pages={719-746},
  year={2005}
}

@article{Wagner2006,
  title={Rheo-SANS investigation of acicular-precipitated calcium carbonate colloidal suspensions through the shear thickening transition},
  author={Egres, R. G. and Nettesheim, F. and Wagner, N. J.},
  journal={J. Rheol.},
  volume={50},
  pages={685-709},
  year={2006}
}

@article{Wagner2009,
  title={Directed self-assembly of suspensions by large amplitude oscillatory shear flow},
  author={McMullan, J. M. and Wagner, N. J.},
  journal={J. Rheol.},
  volume={53},
  pages={575-588},
  year={2009}
}

@article{Wagner2009_2,
  title={Rheology and spatially resolved structure of cetyltrimethylammonium bromide wormlike micelles through the shear banding transition},
  author={Helgeson, M. E. and Vasquez, P. A. and Kaler, E. W. and Wagner, N. J.},
  journal={J. Rheol.},
  volume={53},
  pages={727-756},
  year={2009}
}

@article{Liberatore2006,
  title={Spatially resolved small-angle neutron scattering in the 1-2 plane: A study of shear-induced phase-separating wormlike micelles},
  author={Liberatore, M. W. and Nettesheim, F. and Wagner, N. J. and Porcar, L.},
  journal={Phys. Rev. E},
  volume={73},
  pages={020504},
  year={2006},
  publisher={APS}
}

@article{Liberatore2009,
  title={Microstructure and shear rheology of entangled wormlike micelles in solution},
  author={Liberatore, M. W. and Nettesheim, F. and Vasquez, P. A. and Helgeson, M. E. and Wagner, N. J. and Kaler, E. W. and Cook, L. P. and Porcar, L. and Hu, Y. T.},
  journal={J. Rheol.},
  volume={53},
  pages={441--458},
  year={2009},
  publisher={AIP Publishing}
}

@article{Wagner2016,
  title={Understanding steady and dynamic shear banding in a model wormlike micellar solution},
  author={Calabrese, M. A. and Rogers, S. A. and Porcar, L. and Wagner, N. J.},
  journal={J. Rheol.},
  volume={60},
  pages={1001-1017},
  year={2016}
}

@article{Helgeson2016,
  title={Shear-induced clustering of Brownian colloids in associative polymer networks at moderate Péclet number},
  author={Kim, J. and Helgeson, M. E.},
  journal={Phys. Rev. Fluids},
  volume={1},
  pages={043302},
  year={2016}
}

@article{Porcar2022,
  title={Shear-induced nanostructural changes in micelles formed by sugar-based surfactants with varied anomeric configuration},
  author={Larsson, J. and Williams, A. P. and Wahlgren, M. and Porcar, L. and Ulvenlund, S. and Nylander, T. and Tabor, R. F. and Sanchez-Fernandez, A.},
  journal={J. Colloid Interface Sci.},
  volume={606},
  pages={328–336},
  year={2022}
}

@article{Leadbetter1979,
  title={Distribution functions in three liquid crystals from X-ray diffraction measurements},
  author={Leadbetter, A. J. and Norris, E. K.},
  journal={Mol. Phys.},
  volume={38},
  pages={669-686},
  year={1979}
}

@article{Leadbetter1984,
  title={A correlation between short range smectic-like ordering and the elastic constants of nematic liquid crystals},
  author={Bradshaw, M. J. and Raynes, E. P. and Fedak, I. and Leadbetter, A. J.},
  journal={J. Physique},
  volume={45},
  pages={157-162},
  year={1984}
}

@article{Caspar1988,
  title={Orientational Distribution Function in Nematic Tobacco-Mosaic-Virus Liquid Crystals Measured by X-Ray Diffraction},
  author={Oldenbourg, R. and Wen, X. and Meyer, R. B. and Caspar, D. L. D.},
  journal={Phys. Rev. Lett.},
  volume={61},
  pages={1851-1854},
  year={1988}
}

@article{Deutsch1991,
  title={Orientational order determination in liquid crystals by x-ray diffraction},
  author={Deutsch, M.},
  journal={Phys. Rev. A},
  volume={44},
  pages={8264-8270},
  year={1991}
}

@article{Caspar1993,
  title={Angular correlations and the isotropic-nematic phase transition in suspensions of tobacco mosaic virus},
  author={Fraden, S. and Maret, G. and Caspar, D. L. D.},
  journal={Phys. Rev. E},
  volume={48},
  pages={2816-2837},
  year={1993}
}

@article{Mochrie2003,
  title={Measuring the nematic order of suspensions of colloidal fd virus by x-ray diffraction and optical birefringence},
  author={Purdy, K. R. and Dogic, Z. and Fraden, S. and Rühm, A. and Lurio, L. and Mochrie, S. G. J.},
  journal={Phys. Rev. E},
  volume={67},
  pages={031708},
  year={2003}
}

@article{Heiney2006,
  title={Structure of nematic liquid crystalline elastomers under uniaxial deformation},
  author={Zhang, F. and Heiney, P. A. and Srinivasan, A. and Naciri, J. and Ratna, B.},
  journal={Phys. Rev. E},
  volume={73},
  pages={021701},
  year={2006}
}

@article{Agra-Kooijman2018,
  title={The integrals determining orientational order in liquid crystals by x-ray diffraction revisited},
  author={Agra-Kooijman, D. M. and Fisch, M. R. and Kumar, S.},
  journal={Liq. Cryst.},
  volume={45},
  pages={1},
  year={2018}
}

@article{Sims2019,
  title={Considerations in the determination of orientational order parameters from X-ray scattering experiments},
  author={Sims, M. T. and Abbott, L. C. and Richardson, R. M. and Goodby J. W. and Moore, J. N.},
  journal={Liq. Cryst.},
  volume={46},
  pages={1-7},
  year={2019}
}

@article{Huang2017,
  title={Reconstruction of three-dimensional anisotropic structure from small-angle scattering experiments},
  author={Huang, G.-R. and Wang, Y. and Wu, B. and Wang, Z. and Do, C. Smith, G. S. and Bras, W. and Porcar, L. and Falus, P. and Chen, W.-R.},
  journal={Phys. Rev. E},
  volume={96},
  pages={022612},
  year={2017}
}

@article{Wanger2014SM,
  title={Spatiotemporal stress and structure evolution in dynamically sheared polymer-like micellar solutions},
  author={Gurnon, A. K. and Lopez-Barron, C. R. and Eberle, A. P. R. and Porcar, L. and Wagner, N. J.},
  journal={Soft Matter},
  volume={10},
  pages={2889--2898},
  year={2014}
}

@article{Vainio2014,
  title={Orientation distribution of vertically aligned multiwalled carbon nanotubes},
  author={Vainio, U. and Schnoor, T. I. W. and Koyiloth Vayalil, S. and Schulte, K. and M\"uller, M. and Lilleodden, E. T.},
  journal={J. Phys. Chem. C},
  volume={118},
  pages={9507--9513},
  year={2014},
  publisher={ACS Publications}
}

@article{Zhou2004,
  title={Single wall carbon nanotube fibers extruded from super-acid suspensions: Preferred orientation, electrical, and thermal transport},
  author={Zhou, W. and Vavro, J. and Guthy, C. and Winey, K. I. and Fischer, J. E. and Ericson, L. M. and Ramesh, S. and Saini, R. and Davis, V. A. and Kittrell, C. and Pasquali, M. and Hauge, R. H. and Smalley, R. E.},
  journal={J. Appl. Phys.},
  volume={95},
  pages={649--655},
  year={2004},
  publisher={American Institute of Physics}
}

@article{Wang2006,
  title={Order in vertically aligned carbon nanotube arrays},
  author={Wang, H. and Xu, Z. and Eres, G.},
  journal={Appl. Phys. Letts.},
  volume={88},
  pages={213111},
  year={2006},
  publisher={AIP Publishing}
}

@article{Ch2011,
  title={Quantitative characterization of vertically aligned multi-walled carbon nanotube arrays using small angle x-ray scattering},
  author={Ch Das, N. and Yang, K. and Liu, Y. and Sokol, P. E. and Wang, Z. and Wang, H.},
  journal={J. Nanosci. Nanotechnol.},
  volume={11},
  pages={4995--5000},
  year={2011},
  publisher={American Scientific Publishers}
}

@article{Hwang2000,
  title={Polarized spectroscopy of aligned single-wall carbon nanotubes},
  author={Hwang, J. and Gommans, H. H. and Ugawa, A. and Tashiro, H. and Haggenmueller, R. and Winey, K. I. and Fischer, J. E. and Tanner, D. B. and Rinzler, A. G.},
  journal={Phys. Rev. B},
  volume={62},
  pages={R13310},
  year={2000},
  publisher={APS}
}

@article{HuangJPCL,
  title={Determining Gyration Tensor of Orienting Macromolecules through Their Scattering Signature},
  author={Huang, G.-R. and Wang, Y. and Do, C. and Porcar, L. and Shinohara, Y. and Egami, T. and Chen, W.-R.},
  journal={J. Phys. Chem. Lett.},
  volume={10},
  pages={3978--3984},
  year={2019}
}

@incollection{Lindner1991,
  author    = {Lindner, P.},
  title     = {Steady-state SANS Experiments under External Constraints: The Example of Dilute Polymer Solutons in Laminar Shear Flow},
  publisher = {North-Holland},
  address={Amsterdam},
  year={1991},
  booktitle = {Neutron, X-ray and Light Scattering: Introduction to an Investigative Tool for Colloidal and Polymer Systems, Ed. by P. Lindner, Th. Zemb}
}

@incollection{Lindner2002,
  author    = {Lindner, P.},
  title     = {Scattering Experiments under External Constraints: SANS and Shear Flow},
  publisher = {North-Holland},
  address={Amsterdam},
  year={2002},
  booktitle = {Neutrons, X-rays and Light: Scattering Methods Applied to Soft Condensed Matter, Ed. by P. Lindner, Th. Zemb}
}

@incollection{Lindner2011,
  author    = {Lindner, P. and Schweins, R. and Campell, R. A.},
  title     = {Sample Environment: Soft Matter Sample Environment for Small-Angle Neutron Scattering and Neutron Reflectometry},
  publisher = {John Wiley \& Sons},
  address={Hoboken},
  year={2011},
  booktitle = {Neutrons in Soft Matter, Ed. by T. Imae, T. Kanaya, M. Furusaka, N. Torikai}
}

@article{Martel2021,
  title={Small-angle X-ray and neutron scattering},
  author={Jeffries, C. M. and Ilavsky, J. and Martel, A. and Hinrichs, S. and Meyer, A. and Pedersen, J. S. and Sokolova, A. V. and Svergun, D. I.},
  journal={Nat. Rev. Methods Primers},
  volume={1},
  pages={70},
  year={2021}
}

@article{Gilbert2024,
  title={Advances in sample environments for neutron scattering for colloid and interface science},
  author={Le Brun, A.P. and Gilbert, E. P.},
  journal={Adv. Colloid Interface Sci.},
  volume={327},
  pages={103141},
  year={2024}
}

@article{Solomon2005,
  title={Flow-induced structure in colloidal suspensions},
  author={Vermant, J. and Solomon, M. J.},
  journal={J. Phys.: Condens. Matter},
  volume={17},
  pages={R187–R216},
  year={2005}
}

@article{Tung2024,
  title={Unveiling mesoscopic structures in distorted lamellar phases through deep learning-based small angle neutron scattering analysis},
  author={Tung, C.-H. and Hsiao, Y.-J. and Chen, H.-L. and Huang, G.-R. and Porcar, L. and Chang, M.-C. Carrillo, J.-M. and Wang, Y. and Sumpter, B. G. and Shinohara, Y. and Taylor, J. and Do, C. and Chen, W.-R.},
  journal={J. Colloid Interface Sci.},
  volume={659},
  pages={739-750},
  year={2024}
}

@article{Tung2022,
  title={Small angle scattering of diblock copolymers profiled by machine learning},
  author={Tung, C.-H. and Chang, S.-Y. and Chen, H.-L. and Wang, Y. and Hong, K. and Carrillo, J. M. and Sumpter, B. G. and Shinohara, Y. and Do, C. and Chen, W.-R.},
  journal={J. Chem. Phys.},
  volume={156},
  pages={131101},
  year={2022}
}

@article{Chang2022,
  title={A machine learning inversion scheme for determining interaction from scattering},
  author={Chang, M.-C. and Tung, C.-H. and Chang, S.-Y. and Carrillo, J.-M. and Wang, Y. and Sumpter, B.G. and Huang, G.-R. and Do, C. and Chen, W.-R.},
  journal={Commun. Phys.},
  volume={5},
  pages={46},
  year={2022}
}

@article{Friedel1913,
  title={Sur les symétries cristallines que peut révéler la diffraction des rayons Röntgen},
  author={Friedel, G.},
  journal={Comptes Rendus},
  volume={157},
  pages={1533–1536},
  year={1913}
}

@article{PorcarJoVE,
  title={Measuring Material Microstructure Under Flow Using 1-2 Plane Flow-Small Angle Neutron Scattering},
  author={Gurnon, A. K. and Godfrin, P. D. and Wagner, N. J. and Eberle, A. P. R. Butler, P. and Porcar, L.},
  journal={J. Vis. Exp.},
  volume={84},
  pages={51068},
  year={2014}
}

@article{Odijk1986,
  title={Theory of lyotropic polymer liquid crystals},
  author={Odijk, T.},
  journal={Macromolecules},
  volume={19},
  pages={2313-2329},
  year={1986}
}

@article{MS1958,
  title={Eine einfache molekulare Theorie des nematischen kristallinflüssigen Zustandes},
  author={Maier, W. and Saupe, A.},
  journal={Z. Naturforsch. A},
  volume={13},
  pages={564-566},
  year={1958}
}

@article{HJL1972,
  title={Molecular field treatment of nematic liquid crystals},
  author={Maier, W. and Saupe, A.},
  journal={J. Chem Soc., Faraday Trans. II},
  volume={58},
  pages={1031-1044},
  year={1972}
}

@article{Faber1972,
  title={A continuum theory of disorder in nematic liquid crystals},
  author={Faber, T. E.},
  journal={Proc. R. Soc. Lond. A},
  volume={353},
  pages={247-259},
  year={1977}
}

@article{Stephen1974,
  title={Physics of liquid crystals},
  author={Stephen, M. J. and Straley, J. P.},
  journal={Rev. Mod. Phys.},
  volume={46},
  pages={617-704},
  year={1974}
}

@article{Singh1991,
  title={Density-functional theory of freezing and properties of the ordered phase},
  author={Singh, Y.},
  journal={Phys. Rep.},
  volume={207},
  pages={351-444},
  year={1991}
}

@article{Lekkerkerker1992,
  title={Phase transitions in lyotropic colloidal and polymer liquid crystals},
  author={Vroege, G. J. and Lekkerkerker, H. N. W.},
  journal={Phys. Rep.},
  volume={55},
  pages={1241-1310},
  year={1992}
}

@article{Singh2000,
  title={Phase transitions in liquid crystals},
  author={Singh, S.},
  journal={Phys. Rep.},
  volume={324},
  pages={107-269},
  year={2000}
}

@article{Stokes2020,
  title={A review of nanocrystalline cellulose suspensions: Rheology, liquid crystal ordering and colloidal phase behaviour},
  author={Xu, Y. and Atrens, A. and Stokes, J. R.},
  journal={Adv. Colloid Interface Sci.},
  volume={275},
  pages={102076},
  year={2020}
}

@book{Kardar,
  author = {Kardar, M.},
  title  =  {Statistical Physics of Particles},
  publisher = {Cambridge University Press},
  address={Cambridge},
  year   = {2007}       
}

@book{Oswald,
  author = {Oswald, P.},
  title  =  {Rheophysics: The Deformation and Flow of Matter},
  publisher = {Cambridge University Press},
  address={Cambridge},
  year   = {2009}       
}

@book{Arfken,
  author = {Arfken, G. B. and Weber, H. J. and Harris, F. E.},
  title  =  {Mathematical Methods for Physicists: A Comprehensive Guide},
  publisher = {Academic Press},
  address={Cambridge},
  year   = {2012}       
}

@article{Murphy2020,
  title={Capillary RheoSANS: measuring the rheology and nanostructure of complex fluids at high shear rates},
  author={Murphy, R. P. and Riedel, Z. W. and Nakatani, M. A. and Salipante, P. F. and Weston, J. S. and Hudson, S. D. and Weigandt, K. M.},
  journal={Soft Matter},
  volume={16},
  pages={6285-6293},
  year={2020}
}

@book{Dresselhaus,
  author = {Dresselhaus, M. S. S. and Dresselhaus, G. and Jorio, A.},
  title  =  {Group Theory: Application to the Physics of Condensed Matter},
  publisher = {Springer},
  address={Berlin},
  year   = {2008}       
}

@book{Goldstein,
  author = {Goldstein, H. and Poole, C. and Safko, J.},
  title  =  {Classical Mechanics},
  publisher = {Pearson},
  address={London},
  year   = {2001}       
}

@book{Morawiec,
  author = {Morawiec, A.},
  title  =  {Orientations and Rotations: Computations in Crystallographic Textures},
  publisher = {Springer-Verlag},
  address={Berlin},
  year   = {2004}       
}

@book{Roman,
  author = {Roman, S.},
  title  =  {Advanced Linear Algebra},
  publisher = {Springer-Verlag},
  address={Berlin},
  year   = {2008}       
}

@book{Strang,
  author = {Strang, G.},
  title  =  {Introduction to Linear Algebra},
  publisher = {Wellesley-Cambridge Press},
  address={Wellesley},
  year   = {2016}       
}

@book{Marshall,
  author = {Marshall, W. and Lovesey, S. W.},
  title  =  {Theory of Thermal Neutron Scattering},
  publisher = {Clarendon Press},
  address={Oxford},
  year   = {1971}       
}

@article{Euler,
  title={FORMVLAE GENERALES PRO TRANSLATIONE QVACVNQVE CORPORVM RIGIDORVM},
  author={Evlero, A. L.},
  journal={Novi Comment. Acad. Sci. Imp. Petropol},
  volume={20},
  pages={189-207},
  year={1776}
}

@article{Bonse-Hart,
  title={Small Angle X-Ray Scattering by Spherical Particles of Polystyrene and Polyvinyltoluene},
  author={Bonse, U. and Hart, M.},
  journal={Z. Phys.},
  volume={189},
  pages={151-162},
  year={1966}
}

@article{Burdette-Trofimov1,
  title={Understanding Binder–Silicon Interactions during Slurry Processing},
  author={Burdette-Trofimov, M. K. and Armstrong, B. L. and Rogers, A. M. and Heroux, L. and Doucet, M. and Yang, G. and Phillip, N. D. and Kidder, M. K. and Veith, G. M.},
  journal={J. Chem. Phys. C},
  volume={124},
  pages={13479–13494},
  year={2020}
}

@article{Burdette-Trofimov2,
  title={Probing Clustering Dynamics between Silicon and PAA or LiPAA Slurries under Processing Conditions},
  author={Burdette-Trofimov, M. K. and Armstrong, B. L. and Murphy, R. P. and Heroux, L. and Doucet, M. and Rogers, A. M. and Veith, G. M.},
  journal={ACS Appl. Polym. Mater.},
  volume={3},
  pages={2447–2460},
  year={2021}
}

@article{Burdette-Trofimov3,
  title={Structure and dynamics of small polyimide oligomers with silicon as a function of aging},
  author={Burdette-Trofimov, M. K. and Armstrong, B. L. and Heroux, L. and Doucet, M. and Sacci, R. L. and Veith, G. M.},
  journal={Soft Matter},
  volume={17},
  pages={7729-7742},
  year={2021}
}

@article{Burdette-Trofimov4,
  title={Understanding the Solution Dynamics and Binding of a PVDF Binder with Silicon, Graphite, and NMC Materials and the Influence on Cycling Performance},
  author={Burdette-Trofimov, M. K. and Armstrong, B. L. and Korkosz, R. J. and Tyler, J. L. and McAuliffe, R.D. and Heroux, L. and Doucet, M. and Hoelzer, D. T. and Kanbargi, N. and Naskar, A. K. and Veith, G. M.},
  journal={ACS Appl. Mater. Interfaces},
  volume={14},
  pages={23322-23331},
  year={2022}
}

@article{Burdette-Trofimov5,
  title={Competitive adsorption within electrode slurries and impact on cell fabrication and performance},
  author={Burdette-Trofimov, M. K. and Armstrong, B. L. and Heroux, L. and Doucet, M. and Marquez Rossy, A. E. and Hoelzer, D. T. and Kanbargi, N. and Naskar, A. K. and Veith, G. M.},
  journal={J. Power Sources},
  volume={520},
  pages={230914},
  year={2022}
}

@article{DPD,
  title={Dissipative particle dynamics simulations in colloid and Interface science: a review},
  author={Santo, K. P. and Neimark, A. V.},
  journal={Adv. Colloid Interface Sci.},
  volume={298},
  pages={102545},
  year={2021}
}

@article{Beta1,
  title={Analysis of small angle neutron scattering spectra from polydisperse interacting colloids},
  author={Kotlarchyk, M. and Chen, S.-H.},
  journal={J. Chem. Phys.},
  volume={79},
  pages={2461-2469},
  year={1983}
}

@article{Beta2,
  title={Determination of micelle structure and charge by neutron small-angle scattering},
  author={Hayter, J. B. and Penfold, J.},
  journal={Colloid Polym. Sci.},
  volume={261},
  pages={1022-1030},
  year={1983}
}

@article{Lam2018Scaling,
  author  = {Lam, C. N. and Xu, W.-S. and Chen, W.-R. and Wang, Z. and Stanley, C. B. and Carrillo, J.-M. Y. and Uhrig, D. and Wang, W. and Hong, K. and Liu, Y. and Porcar, L. and Do, C. and Smith, G. S. and Sumpter, B. G. and Wang, Y.},
  title   = {Scaling Behavior of Anisotropy Relaxation in Deformed Polymers},
  journal = {Phys. Rev. Lett.},
  volume  = {121},
  number  = {11},
  pages   = {117801},
  doi     = {10.1103/PhysRevLett.121.117801},
  url     = {https://link.aps.org/doi/10.1103/PhysRevLett.121.117801},
  year    = {2018},
  type    = {Journal Article}
}

@article{lam2023quantifying,
  title={Quantifying molecular deformation in polymer melts by a generalized Zimm plot approach},
  author={Lam, Christopher N and He, Lilin and Do, Changwoo and Chen, W-R and Wang, Weiyu and Hong, Kunlun and Wang, Yangyang},
  journal={J. Appl. Crystallogr.},
  volume={56},
  number={4},
  pages={1168-1179},
  year={2023},
  publisher={International Union of Crystallography}
}

@article{sun2021molecular,
  title     = {Molecular view on mechanical reinforcement in polymer nanocomposites},
  author    = {Sun, Ruikun and Melton, Matthew and Safaie, Niloofar and Ferrier Jr, Robert C and Cheng, Shiwang and Liu, Yun and Zuo, Xiaobing and Wang, Yangyang},
  journal   = {Phys. Rev. Lett.},
  volume    = {126},
  number    = {11},
  pages     = {117801},
  year      = {2021},
  publisher = {APS}
}

@article{wang2017fingerprinting,
  title     = {Fingerprinting molecular relaxation in deformed polymers},
  author    = {Wang, Zhe and Lam, Christopher N and Chen, Wei-Ren and Wang, Weiyu and Liu, Jianning and Liu, Yun and Porcar, Lionel and Stanley, Christopher B and Zhao, Zhichen and Hong, Kunlun and Yangyang Wang},
  journal   = {Phys. Rev. X},
  volume    = {7},
  number    = {3},
  pages     = {031003},
  year      = {2017},
  publisher = {APS}
}

@article{wang2020quantitative,
  title     = {Quantitative examination of a fundamental assumption in small-angle neutron scattering studies of deformed polymer melts},
  author    = {Wang, Yangyang and Wang, Weiyu and Hong, Kunlun and Do, Changwoo and Chen, Wei-Ren},
  journal   = {Polymer},
  volume    = {204},
  pages     = {122698},
  year      = {2020},
  publisher = {Elsevier}
}


\clearpage

\end{document}